\documentclass[11pt,a4paper]{article}
\pdfoutput=1

\usepackage{lineno}

\usepackage{jinstpub}
\usepackage{graphicx}
\usepackage{subfig}
\usepackage{diffcoeff}
\usepackage{siunitx}
\DeclareSIUnit\ppm{ppm}
\usepackage[version=3]{mhchem}
\usepackage{float}
\usepackage{subfig}
\usepackage{comment}
\usepackage{hyperref}
\usepackage{orcidlink}

\title{Doping liquid argon with xenon in ProtoDUNE Single-Phase: effects on scintillation light}

\date{Draft n.1 (13 March 2022)}

\collaboration{The DUNE Collaboration}

\affiliation[0]{Abilene Christian University, Abilene, TX 79601, USA}
\affiliation[1]{University of Albany, SUNY, Albany, NY 12222, USA}
\affiliation[2]{University of Amsterdam, NL-1098 XG Amsterdam, The Netherlands}
\affiliation[3]{Antalya Bilim University, 07190 D{\"o}{\c{s}}emealt{\i}/Antalya, Turkey}
\affiliation[4]{University of Antananarivo, Antananarivo 101, Madagascar}
\affiliation[5]{University of Antioquia, Medell{\'\i}n, Colombia}
\affiliation[6]{Universidad Antonio Nari{\~n}o, Bogot{\'a}, Colombia}
\affiliation[7]{Argonne National Laboratory, Argonne, IL 60439, USA}
\affiliation[8]{University of Arizona, Tucson, AZ 85721, USA}
\affiliation[9]{Universidad Nacional de Asunci{\'o}n, San Lorenzo, Paraguay}
\affiliation[10]{University of Athens, Zografou GR 157 84, Greece}
\affiliation[11]{Universidad del Atl{\'a}ntico, Barranquilla, Atl{\'a}ntico, Colombia}
\affiliation[12]{Augustana University, Sioux Falls, SD 57197, USA}
\affiliation[13]{University of Bern, CH-3012 Bern, Switzerland}
\affiliation[14]{Beykent University, Istanbul, Turkey}
\affiliation[15]{University of Birmingham, Birmingham B15 2TT, United Kingdom}
\affiliation[16]{Universit{\`a} del Bologna, 40127 Bologna, Italy}
\affiliation[17]{Boston University, Boston, MA 02215, USA}
\affiliation[18]{University of Bristol, Bristol BS8 1TL, United Kingdom}
\affiliation[19]{Brookhaven National Laboratory, Upton, NY 11973, USA}
\affiliation[20]{University of Bucharest, Bucharest, Romania}
\affiliation[21]{University of California Berkeley, Berkeley, CA 94720, USA}
\affiliation[22]{University of California Davis, Davis, CA 95616, USA}
\affiliation[23]{University of California Irvine, Irvine, CA 92697, USA}
\affiliation[24]{University of California Los Angeles, Los Angeles, CA 90095, USA}
\affiliation[25]{University of California Riverside, Riverside CA 92521, USA}
\affiliation[26]{University of California Santa Barbara, Santa Barbara, California 93106 USA}
\affiliation[27]{California Institute of Technology, Pasadena, CA 91125, USA}
\affiliation[28]{University of Cambridge, Cambridge CB3 0HE, United Kingdom}
\affiliation[29]{Universidade Estadual de Campinas, Campinas - SP, 13083-970, Brazil}
\affiliation[30]{Universit{\`a} di Catania, 2 - 95131 Catania, Italy}
\affiliation[31]{Universidad Cat{\'o}lica del Norte, Antofagasta, Chile}
\affiliation[32]{Centro Brasileiro de Pesquisas F\'isicas, Rio de Janeiro, RJ 22290-180, Brazil}
\affiliation[33]{IRFU, CEA, Universit{\'e} Paris-Saclay, F-91191 Gif-sur-Yvette, France}
\affiliation[34]{CERN, The European Organization for Nuclear Research, 1211 Meyrin, Switzerland}
\affiliation[35]{Institute of Particle and Nuclear Physics of the Faculty of Mathematics and Physics of the Charles University, 180 00 Prague 8, Czech Republic }
\affiliation[36]{University of Chicago, Chicago, IL 60637, USA}
\affiliation[37]{Chung-Ang University, Seoul 06974, South Korea}
\affiliation[38]{CIEMAT, Centro de Investigaciones Energ{\'e}ticas, Medioambientales y Tecnol{\'o}gicas, E-28040 Madrid, Spain}
\affiliation[39]{University of Cincinnati, Cincinnati, OH 45221, USA}
\affiliation[40]{Centro de Investigaci{\'o}n y de Estudios Avanzados del Instituto Polit{\'e}cnico Nacional (Cinvestav), Mexico City, Mexico}
\affiliation[41]{Universidad de Colima, Colima, Mexico}
\affiliation[42]{University of Colorado Boulder, Boulder, CO 80309, USA}
\affiliation[43]{Colorado State University, Fort Collins, CO 80523, USA}
\affiliation[44]{Columbia University, New York, NY 10027, USA}
\affiliation[45]{Comisi{\'o}n Nacional de Investigaci{\'o}n y Desarrollo Aeroespacial, Lima, Peru}
\affiliation[46]{Centro de Tecnologia da Informa{\c{c}}ao Renato Archer, Amarais - Campinas, SP - CEP 13069-901, Brazil}
\affiliation[47]{Central University of South Bihar, Gaya, 824236, India }
\affiliation[48]{Institute of Physics, Czech Academy of Sciences, 182 00 Prague 8, Czech Republic}
\affiliation[49]{Czech Technical University, 115 19 Prague 1, Czech Republic}
\affiliation[50]{Laboratoire d{\textquoteright}Annecy de Physique des Particules, Universit{\'e} Savoie Mont Blanc, CNRS, LAPP-IN2P3, 74000 Annecy, France}
\affiliation[51]{Daresbury Laboratory, Cheshire WA4 4AD, United Kingdom}
\affiliation[52]{Dordt University, Sioux Center, IA 51250, USA}
\affiliation[53]{Drexel University, Philadelphia, PA 19104, USA}
\affiliation[54]{Duke University, Durham, NC 27708, USA}
\affiliation[55]{Durham University, Durham DH1 3LE, United Kingdom}
\affiliation[56]{University of Edinburgh, Edinburgh EH8 9YL, United Kingdom}
\affiliation[57]{Universidad EIA, Envigado, Antioquia, Colombia}
\affiliation[58]{Erciyes University, Kayseri, Turkey}
\affiliation[59]{ETH Zurich, Zurich, Switzerland}
\affiliation[60]{E{\"o}tv{\"o}s Lor{\'a}nd University, 1053 Budapest, Hungary}
\affiliation[61]{Faculdade de Ci{\^e}ncias da Universidade de Lisboa - FCUL, 1749-016 Lisboa, Portugal}
\affiliation[62]{Universidade Federal de Alfenas, Po{\c{c}}os de Caldas - MG, 37715-400, Brazil}
\affiliation[63]{Universidade Federal de Goias, Goiania, GO 74690-900, Brazil}
\affiliation[64]{Universidade Federal do ABC, Santo Andr{\'e} - SP, 09210-580, Brazil}
\affiliation[65]{Universidade Federal do Rio de Janeiro,  Rio de Janeiro - RJ, 21941-901, Brazil}
\affiliation[66]{Fermi National Accelerator Laboratory, Batavia, IL 60510, USA}
\affiliation[67]{University of Ferrara, Ferrara, Italy}
\affiliation[68]{University of Florida, Gainesville, FL 32611-8440, USA}
\affiliation[69]{Florida State University, Tallahassee, FL, 32306 USA}
\affiliation[70]{Fluminense Federal University, 9 Icara{\'\i} Niter{\'o}i - RJ, 24220-900, Brazil }
\affiliation[71]{Universit{\`a} degli Studi di Genova, Genova, Italy}
\affiliation[72]{Georgian Technical University, Tbilisi, Georgia}
\affiliation[73]{University of Granada {\&} CAFPE, 18002 Granada, Spain}
\affiliation[74]{Gran Sasso Science Institute, L'Aquila, Italy}
\affiliation[75]{Laboratori Nazionali del Gran Sasso, L'Aquila AQ, Italy}
\affiliation[76]{University Grenoble Alpes, CNRS, Grenoble INP, LPSC-IN2P3, 38000 Grenoble, France}
\affiliation[77]{Universidad de Guanajuato, Guanajuato, C.P. 37000, Mexico}
\affiliation[78]{Harish-Chandra Research Institute, Jhunsi, Allahabad 211 019, India}
\affiliation[79]{University of Hawaii, Honolulu, HI 96822, USA}
\affiliation[80]{Hong Kong University of Science and Technology, Kowloon, Hong Kong, China}
\affiliation[81]{University of Houston, Houston, TX 77204, USA}
\affiliation[82]{University of  Hyderabad, Gachibowli, Hyderabad - 500 046, India}
\affiliation[83]{Idaho State University, Pocatello, ID 83209, USA}
\affiliation[84]{Instituto de F{\'\i}sica Corpuscular, CSIC and Universitat de Val{\`e}ncia, 46980 Paterna, Valencia, Spain}
\affiliation[85]{Instituto Galego de F{\'\i}sica de Altas Enerx{\'\i}as, University of Santiago de Compostela, Santiago de Compostela, 15782, Spain}
\affiliation[86]{Indian Institute of Technology Kanpur, Uttar Pradesh 208016, India}
\affiliation[87]{Illinois Institute of Technology, Chicago, IL 60616, USA}
\affiliation[88]{Imperial College of Science Technology and Medicine, London SW7 2BZ, United Kingdom}
\affiliation[89]{Indian Institute of Technology Guwahati, Guwahati, 781 039, India}
\affiliation[90]{Indian Institute of Technology Hyderabad, Hyderabad, 502285, India}
\affiliation[91]{Indiana University, Bloomington, IN 47405, USA}
\affiliation[92]{Istituto Nazionale di Fisica Nucleare Sezione di Bologna, 40127 Bologna BO, Italy}
\affiliation[93]{Istituto Nazionale di Fisica Nucleare Sezione di Catania, I-95123 Catania, Italy}
\affiliation[94]{Istituto Nazionale di Fisica Nucleare Sezione di Ferrara, I-44122 Ferrara, Italy}
\affiliation[95]{Istituto Nazionale di Fisica Nucleare Laboratori Nazionali di Frascati, Frascati, Roma, Italy}
\affiliation[96]{Istituto Nazionale di Fisica Nucleare Sezione di Genova, 16146 Genova GE, Italy}
\affiliation[97]{Istituto Nazionale di Fisica Nucleare Sezione di Lecce, 73100 - Lecce, Italy}
\affiliation[98]{Istituto Nazionale di Fisica Nucleare Sezione di Milano Bicocca, 3 - I-20126 Milano, Italy}
\affiliation[99]{Istituto Nazionale di Fisica Nucleare Sezione di Milano, 20133 Milano, Italy}
\affiliation[100]{Istituto Nazionale di Fisica Nucleare Sezione di Napoli, I-80126 Napoli, Italy}
\affiliation[101]{Istituto Nazionale di Fisica Nucleare Sezione di Padova, 35131 Padova, Italy}
\affiliation[102]{Istituto Nazionale di Fisica Nucleare Sezione di Pavia,  I-27100 Pavia, Italy}
\affiliation[103]{Istituto Nazionale di Fisica Nucleare Laboratori Nazionali di Pisa, Pisa PI, Italy}
\affiliation[104]{Istituto Nazionale di Fisica Nucleare Sezione di Roma, 00185 Roma RM, Italy}
\affiliation[105]{Istituto Nazionale di Fisica Nucleare Laboratori Nazionali del Sud, 95123 Catania, Italy}
\affiliation[106]{Universidad Nacional de Ingenier{\'\i}a, Lima 25, Per{\'u}}
\affiliation[108]{University of Insubria, Via Ravasi, 2, 21100 Varese VA, Italy}
\affiliation[109]{University of Iowa, Iowa City, IA 52242, USA}
\affiliation[110]{Iowa State University, Ames, Iowa 50011, USA}
\affiliation[111]{Institut de Physique des 2 Infinis de Lyon, 69622 Villeurbanne, France}
\affiliation[112]{Institute for Research in Fundamental Sciences, Tehran, Iran}
\affiliation[113]{Instituto Superior T{\'e}cnico - IST, Universidade de Lisboa, Portugal}
\affiliation[114]{Instituto Tecnol{\'o}gico de Aeron{\'a}utica, Sao Jose dos Campos, Brazil}
\affiliation[115]{Iwate University, Morioka, Iwate 020-8551, Japan}
\affiliation[116]{Jackson State University, Jackson, MS 39217, USA}
\affiliation[117]{Jawaharlal Nehru University, New Delhi 110067, India}
\affiliation[118]{Jeonbuk National University, Jeonrabuk-do 54896, South Korea}
\affiliation[120]{Jyv{\"a}skyl{\"a} University, FI-40014 Jyv{\"a}skyl{\"a}, Finland}
\affiliation[121]{Kansas State University, Manhattan, KS 66506, USA}
\affiliation[122]{Kavli Institute for the Physics and Mathematics of the Universe, Kashiwa, Chiba 277-8583, Japan}
\affiliation[123]{High Energy Accelerator Research Organization (KEK), Ibaraki, 305-0801, Japan}
\affiliation[124]{Korea Institute of Science and Technology Information, Daejeon, 34141, South Korea}
\affiliation[125]{National Institute of Technology, Kure College, Hiroshima, 737-8506, Japan}
\affiliation[126]{Taras Shevchenko National University of Kyiv, 01601 Kyiv, Ukraine}
\affiliation[127]{Lancaster University, Lancaster LA1 4YB, United Kingdom}
\affiliation[128]{Lawrence Berkeley National Laboratory, Berkeley, CA 94720, USA}
\affiliation[129]{Laborat{\'o}rio de Instrumenta{\c{c}}{\~a}o e F{\'\i}sica Experimental de Part{\'\i}culas, 1649-003 Lisboa and 3004-516 Coimbra, Portugal}
\affiliation[130]{University of Liverpool, L69 7ZE, Liverpool, United Kingdom}
\affiliation[131]{Los Alamos National Laboratory, Los Alamos, NM 87545, USA}
\affiliation[132]{Louisiana State University, Baton Rouge, LA 70803, USA}
\affiliation[133]{University of Lucknow, Uttar Pradesh 226007, India}
\affiliation[134]{Madrid Autonoma University and IFT UAM/CSIC, 28049 Madrid, Spain}
\affiliation[135]{Johannes Gutenberg-Universit{\"a}t Mainz, 55122 Mainz, Germany}
\affiliation[136]{University of Manchester, Manchester M13 9PL, United Kingdom}
\affiliation[137]{Massachusetts Institute of Technology, Cambridge, MA 02139, USA}
\affiliation[138]{University of Medell{\'\i}n, Medell{\'\i}n, 050026 Colombia }
\affiliation[139]{University of Michigan, Ann Arbor, MI 48109, USA}
\affiliation[140]{Michigan State University, East Lansing, MI 48824, USA}
\affiliation[141]{Universit{\`a} di Milano Bicocca , 20126 Milano, Italy}
\affiliation[142]{Universit{\`a} degli Studi di Milano, I-20133 Milano, Italy}
\affiliation[143]{University of Minnesota Duluth, Duluth, MN 55812, USA}
\affiliation[144]{University of Minnesota Twin Cities, Minneapolis, MN 55455, USA}
\affiliation[145]{University of Mississippi, University, MS 38677 USA}
\affiliation[146]{Universit{\`a} degli Studi di Napoli Federico II , 80138 Napoli NA, Italy}
\affiliation[147]{Nikhef National Institute of Subatomic Physics, 1098 XG Amsterdam, Netherlands}
\affiliation[148]{National Institute of Science Education and Research (NISER), Odisha 752050, India}
\affiliation[149]{University of North Dakota, Grand Forks, ND 58202-8357, USA}
\affiliation[150]{Northern Illinois University, DeKalb, IL 60115, USA}
\affiliation[151]{Northwestern University, Evanston, Il 60208, USA}
\affiliation[152]{University of Notre Dame, Notre Dame, IN 46556, USA}
\affiliation[153]{University of Novi Sad, 21102 Novi Sad, Serbia}
\affiliation[154]{Occidental College, Los Angeles, CA  90041, USA}
\affiliation[155]{Ohio State University, Columbus, OH 43210, USA}
\affiliation[156]{Oregon State University, Corvallis, OR 97331, USA}
\affiliation[157]{University of Oxford, Oxford, OX1 3RH, United Kingdom}
\affiliation[158]{Pacific Northwest National Laboratory, Richland, WA 99352, USA}
\affiliation[159]{Universt{\`a} degli Studi di Padova, I-35131 Padova, Italy}
\affiliation[160]{Panjab University, Chandigarh, 160014, India}
\affiliation[161]{Universit{\'e} Paris-Saclay, CNRS/IN2P3, IJCLab, 91405 Orsay, France}
\affiliation[162]{Universit{\'e} Paris Cit{\'e}, CNRS, Astroparticule et Cosmologie, Paris, France}
\affiliation[163]{University of Parma,  43121 Parma PR, Italy}
\affiliation[164]{Universit{\`a} degli Studi di Pavia, 27100 Pavia PV, Italy}
\affiliation[165]{University of Pennsylvania, Philadelphia, PA 19104, USA}
\affiliation[166]{Pennsylvania State University, University Park, PA 16802, USA}
\affiliation[167]{Physical Research Laboratory, Ahmedabad 380 009, India}
\affiliation[168]{Universit{\`a} di Pisa, I-56127 Pisa, Italy}
\affiliation[169]{University of Pittsburgh, Pittsburgh, PA 15260, USA}
\affiliation[170]{Pontificia Universidad Cat{\'o}lica del Per{\'u}, Lima, Per{\'u}}
\affiliation[171]{University of Puerto Rico, Mayaguez 00681, Puerto Rico, USA}
\affiliation[172]{Punjab Agricultural University, Ludhiana 141004, India}
\affiliation[173]{Queen Mary University of London, London E1 4NS, United Kingdom }
\affiliation[174]{Radboud University, NL-6525 AJ Nijmegen, Netherlands}
\affiliation[175]{Rice University, Houston, TX 77005, USA}
\affiliation[176]{University of Rochester, Rochester, NY 14627, USA}
\affiliation[177]{Royal Holloway College London, London, TW20 0EX, United Kingdom}
\affiliation[178]{Rutgers University, Piscataway, NJ, 08854, USA}
\affiliation[179]{STFC Rutherford Appleton Laboratory, Didcot OX11 0QX, United Kingdom}
\affiliation[180]{Universit{\`a} del Salento, 73100 Lecce, Italy}
\affiliation[181]{Universidad del Magdalena, Santa Marta - Colombia}
\affiliation[182]{Sapienza University of Rome, 00185 Roma RM, Italy}
\affiliation[183]{Universidad Sergio Arboleda, 11022 Bogot{\'a}, Colombia}
\affiliation[184]{University of Sheffield, Sheffield S3 7RH, United Kingdom}
\affiliation[185]{SLAC National Accelerator Laboratory, Menlo Park, CA 94025, USA}
\affiliation[186]{University of South Carolina, Columbia, SC 29208, USA}
\affiliation[187]{South Dakota School of Mines and Technology, Rapid City, SD 57701, USA}
\affiliation[188]{South Dakota State University, Brookings, SD 57007, USA}
\affiliation[189]{Southern Methodist University, Dallas, TX 75275, USA}
\affiliation[190]{Stony Brook University, SUNY, Stony Brook, NY 11794, USA}
\affiliation[191]{Sun Yat-Sen University, Guangzhou, 510275, China}
\affiliation[192]{Sanford Underground Research Facility, Lead, SD, 57754, USA}
\affiliation[193]{University of Sussex, Brighton, BN1 9RH, United Kingdom}
\affiliation[194]{Syracuse University, Syracuse, NY 13244, USA}
\affiliation[195]{Universidade Tecnol{\'o}gica Federal do Paran{\'a}, Curitiba, Brazil}
\affiliation[196]{Tel Aviv University, Tel Aviv-Yafo, Israel}
\affiliation[197]{Texas A{\&}M University, College Station, Texas 77840, USA}
\affiliation[198]{Texas A{\&}M University - Corpus Christi, Corpus Christi, TX 78412, USA}
\affiliation[199]{University of Texas at Arlington, Arlington, TX 76019, USA}
\affiliation[200]{University of Texas at Austin, Austin, TX 78712, USA}
\affiliation[201]{University of Toronto, Toronto, Ontario M5S 1A1, Canada}
\affiliation[202]{Tufts University, Medford, MA 02155, USA}
\affiliation[203]{Universidade Federal de S{\~a}o Paulo, 09913-030, S{\~a}o Paulo, Brazil}
\affiliation[204]{Ulsan National Institute of Science and Technology, Ulsan 689-798, South Korea}
\affiliation[205]{University College London, London, WC1E 6BT, United Kingdom}
\affiliation[206]{Universidad Nacional Mayor de San Marcos, Lima, Peru}
\affiliation[207]{Valley City State University, Valley City, ND 58072, USA}
\affiliation[208]{Virginia Tech, Blacksburg, VA 24060, USA}
\affiliation[209]{University of Warsaw, 02-093 Warsaw, Poland}
\affiliation[210]{University of Warwick, Coventry CV4 7AL, United Kingdom}
\affiliation[211]{Wellesley College, Wellesley, MA 02481, USA}
\affiliation[212]{Wichita State University, Wichita, KS 67260, USA}
\affiliation[213]{William and Mary, Williamsburg, VA 23187, USA}
\affiliation[214]{University of Wisconsin Madison, Madison, WI 53706, USA}
\affiliation[215]{Yale University, New Haven, CT 06520, USA}
\affiliation[216]{Yerevan Institute for Theoretical Physics and Modeling, Yerevan 0036, Armenia}
\affiliation[217]{York University, Toronto M3J 1P3, Canada}

\author[34]{A.~Abed Abud,}
\author[157]{B.~Abi,}
\author[66]{R.~Acciarri,}
\author[11]{M.~A.~Acero,}
\author[195]{M.~R.~Adames,}
\author[72]{G.~Adamov,}
\author[66]{M.~Adamowski,}
\author[19]{D.~Adams,}
\author[18]{M.~Adinolfi,}
\author[29]{C.~Adriano,}
% ...

\emailAdd{andrea.zani@mi.infn.it; ngallice@bnl.gov}

\abstract{Doping of liquid argon TPCs (LArTPCs) with a small concentration of xenon is a technique for light-shifting and facilitates the detection of the liquid argon scintillation light. In this paper, we present the results of the first doping test ever performed in a kiloton-scale LArTPC. From February to May 2020, we carried out this special run in the single-phase DUNE Far Detector prototype (ProtoDUNE-SP) at CERN, featuring 720~t of total liquid argon mass with 410~t of fiducial mass. A 5.4~ppm nitrogen contamination was present during the xenon doping campaign. The goal of the run was to measure the light and charge response of the detector to the addition of xenon, up to a concentration of 18.8~ppm. The main purpose was to test the possibility for reduction of non-uniformities in light collection, caused by deployment of photon detectors only within the anode planes.
Light collection was analysed as a function of the xenon concentration, by using the pre-existing photon detection system (PDS) of ProtoDUNE-SP and an additional smaller set-up installed specifically for this run.
In this paper we first summarize our current understanding of the argon-xenon energy transfer process and the impact of the presence of nitrogen in argon with and without xenon dopant. We then describe the key elements of ProtoDUNE-SP and the injection method deployed. 
Two dedicated photon detectors were able to collect the light produced by xenon and the total light. The ratio of these components was measured to be about 0.65 as 18.8~ppm of xenon were injected.
We performed studies of the collection efficiency as a function of the distance between tracks and light detectors, demonstrating enhanced uniformity of response for the anode-mounted PDS. We also show that xenon doping can substantially recover light losses due to contamination of the liquid argon by nitrogen.}


\begin{document}

\maketitle

\newpage 
\section{Introduction}
\label{sec:intro}

Liquid Argon Time Projection Chambers (LArTPCs,~\cite{Rubbia:117852}) are prominent in contemporary physics for the study of neutrino oscillations and interactions, and the search for rare events, such as dark matter interactions~\cite{icarus_AMERIO2004329,microboone_Acciarri_2017,argoneut_Anderson_2012,AGNES2015456}. This technology has been developed for more than forty years and has reached a level of sophistication such that it is scalable up to multi-kiloton neutrino detectors, such as the Deep Underground Neutrino Experiment (DUNE)~\cite{Abi_2020_V1,Abi_2020_V4}. 

Particles deposit energy in a LArTPC through ionization and excitation. The ionization charge deposited in argon, drifted towards the anode plane under the influence of a uniform electric field, is exploited to perform spatial and calorimetric reconstruction of events.  Liquid argon (LAr) is also a high-performance scintillator. Excitation leads to light emission in the vacuum ultraviolet (VUV) region with a spectrum centered at $\lambda=127$~nm~\cite{Heindl_2010} and a yield of about $4.0 \times 10^4$ ($2.4 \times 10^4$)~photons/MeV at 0~V/cm (500~V/cm) electric field~\cite{Doke_2002}. 
The scintillation light is produced by de-excitation of the singlet 
($\tau_s \simeq 6$ ns) and triplet ($\tau_l \simeq 1.6 \ \mu$s) states of the unstable excited dimer Ar$_2^*$~\cite{DOKE199962}; their ratio depends on the energy loss mechanism and can be used for particle identification by characterizing the profile of the scintillation light pulse, i.e., the pulse shape.

Detecting VUV light in liquid argon is more challenging than with visible light, and it usually requires special materials or coatings on photodetectors; however, the physics advantages are remarkable.
The scintillation light provides the interaction time ($t_0$) with a few nanoseconds precision, allowing reconstruction of the third spatial coordinate in the TPC.
This improves by an order of magnitude (1~cm $\rightarrow$ 1~mm) the localization of the interaction vertex, compared to using the $t_0$ provided by the proton kicker of the neutrino beam~\cite{Abi_2020_V4}.
Light collection is also the main tool to trigger events that are not produced by the beam, such as interactions of neutrinos from core-collapse supernovae.
The amount of scintillation light produced is anti-correlated with the ionization energy loss of the particle, a feature that can be exploited for combined charge-light calorimetry~\cite{PhysRevD.101.012010}. A high light collection efficiency can result in an increased energy resolution, outperforming that from the ionization signal alone, especially for low energy events in the region of few~MeV to few tens of~MeV~\cite{SNdune}.

The DUNE photon detection system (PDS) can be enhanced by doping LAr with xenon at the level of few tens of ppm\footnote{In this paper, unless otherwise specified, the fractional amounts ppm, ppb, ppt (parts per million/billion/trillion) are to be intended as expressing fractions of \textit{mass}.}. 
DUNE is exploring this possibility because the xenon emission can be collected with higher efficiency due to its longer wavelength with respect to argon: 178~nm~\cite{doi:10.1063/1.1695927} instead of 127~nm. Furthermore, as it will be detailed later, the longer Rayleigh scattering length of the xenon photons in LAr~\cite{128lightspeed} should enhance light collection far from the photon detectors~\cite{gamez_rayleigh}. 
Previous literature studies~\cite{Kubota_1982, Suzuki_1993, Hofmann_2013, Wahl_2014, Akimov_2019} have demonstrated the doping procedure in small scale detectors, and sometimes in gas phase. In order to test the feasibility of such an operation in DUNE, which foresees deployment of four underground modules with a total mass of 17~kt each, it is necessary to demonstrate it at an intermediate scale; therefore, a dedicated xenon doping run was performed in 2020 with the 720~t single-phase DUNE Far Detector prototype at CERN (ProtoDUNE Single-Phase, SP)~\cite{pdsp_design, Abi_2020_PDSPperf}, which represents a new milestone in the development of very-large-volume (multi-kt scale) LArTPCs.

For the work described in this paper, the ProtoDUNE-SP PDS was enhanced with the addition of two prototypes of the second generation X-ARAPUCA photon detectors~\cite{Machado_2018}, which is the technology selected for deployment in the first two modules of the DUNE far detector (referred to as FD1 and FD2).

In this paper, we describe the preparation for and the results of the xenon doping run of ProtoDUNE-SP, obtained both with the X-ARAPUCA and the original PDS light collectors. The physics of light production in xenon-doped liquid argon is introduced in section~\ref{sec:xe_doping}; ProtoDUNE-SP and its photon detection system is described in section~\ref{sec:ProtoDUNE-SP}; the xenon doping procedure is detailed in section~\ref{sec:cryostat}. 
The analysis of the data recorded by the X-ARAPUCA is presented in section~\ref{sec:analysis_xArapuca}; the studies performed with the main PDS are shown in section~\ref{sec:analysis_protodune_pds}. Finally, in section~\ref{sec:charge_analysis}, we use tracks reconstructed in the TPC to evaluate the effect of the xenon presence on the charge collection.

\section{Xenon doping of liquid argon}
\label{sec:xe_doping}

Xenon liquefies at 165~K and freezes at 161~K; it is a high-yield scintillator. The average energy needed to produce a scintillation photon in xenon is slightly lower than in argon, for both low- and high-ionization density particles~\cite{DOKE199962}. This results in a slightly higher photon yield $> 4.2 \times 10^4$~photons/MeV, without electric field (to be compared with $4.0 \times 10^4$~photons/MeV for LAr, as mentioned in sec.~\ref{sec:intro}).
Xenon scintillation light is emitted at 178~nm, as compared to 127~nm for argon, and also features two components ($\tau_f = 2-4 ~\si{\nano\second}$ and $\tau_s = 22 -24 ~\si{\nano\second}$) \cite{hitachi,PhysRevD.97.112002} that are both much faster than the argon triplet light.

Xenon has been exploited as a cryogenic liquid in various direct dark matter search experiments~\cite{xenon1t, lux_AKERIB2013111} and in the quest for neutrinoless double beta decay~\cite{PhysRevLett.123.161802} but, due to its low availability and high production cost, its use as the primary component in large-scale neutrino detectors is quite limited. 
However, as already mentioned, there have been several studies of its beneficial light production properties when used as low concentration dopant, limited to relatively small detectors (see for instance~\cite{Kubota_1982, Suzuki_1993, Hofmann_2013, Wahl_2014, Akimov_2019}).
Motivated by these studies, the DUNE Collaboration initiated a program to investigate the possibility of using xenon doping to enhance the photon detection in the massive 17~kt modules. 

\subsection{Doping liquid argon with xenon and its advantages}

Converting liquid argon scintillation light to a longer wavelength has significant advantages in a LArTPC, especially if it can be achieved uniformly throughout the drift volume rather than on the surface of photosensitive devices (as is the case for standard wavelength-shifting coatings).
At the xenon wavelength (178~nm), light detectors with high enough sensitivity are already commercially available.
For example, the photon detection efficiency (PDE) of current-generation silicon photomultipliers (SiPMs) at this wavelength exceeds 15\%~\cite{nexo-fbk,nexo-hpk}. 
This would ensure quite efficient collection of the xenon light while mitigating the need for wavelength-shifting coatings, such as tetra-phenyl butadiene (TPB). However, it can be beneficial, for effective large-area detection of light, to use more elaborate configurations, such as the DUNE X-ARAPUCA light trap.

One of the primary benefits of the longer wavelength for large detectors is that the Rayleigh scattering length ($\mathcal{L}_R$) for 178~nm light in liquid argon  is significantly longer than that for 127~nm light. This is largely due to the strong dependence of $\mathcal{L}_R$ on the wavelength, as shown in~\cite{128lightspeed}. From the reference, we obtain $\mathcal{L}_R = 1 (8.3)$~m for $\lambda = 127 (178)$~nm, respectively.
For DUNE, this will mean a more uniform response to photons reaching the light detectors with less dependence on the distance between the ionizing source and the photon detectors. In particular, this may improve the trigger efficiency for non-beam-related low energy events far from the light detectors.

The faster de-excitation decay time constants of xenon (4~ns and 22~ns) contribute to shorter pulse profiles, with respect to argon: in undoped argon, the singlet-to-triplet ratio for minimum ionizing particles (MIPs) is about 0.3, resulting in a rather slow pulse~\cite{hitachi}. Even considering the convolution of the various processes involved in argon-xenon excitation transfer, one can obtain signals with an overall decay constant of a few hundreds nanoseconds.

These properties of xenon make it an attractive option for DUNE. The main focus of the doping campaign in ProtoDUNE was indeed to evaluate its feasibility and the advantages for the DUNE Far Detectors in a real, large-scale set up. Furthermore, boosting the doping at the level of a few percent could enhance the physics goals of DUNE, making one of the far detector modules a next-generation, neutrinoless double beta decay experiment~\cite{PhysRevD.106.092002}.

\subsection{Mechanism of xenon-doped LAr scintillation}

The xenon concentration levels used in previous experiments ranged from a few~ppm to a few percent, with light shifting effects detectable even at the lower values~\cite{Kubota_1982, Suzuki_1993, Hofmann_2013, Wahl_2014, Akimov_2019}.
According to current models~\cite{Buzulutskov_2017}, xenon atoms in suspension in liquid argon interfere with the light production process that involves the argon excited dimers Ar$_2^*$. These dimers form in two states, a singlet $^1\Sigma^+_u$ characterized by a fast decay constant (6~ns, thus dubbed in the following ``fast component''), and a triplet state $^3\Sigma^+_u$ with a much larger decay time (up to $\sim$1600~ns, ``slow component''\footnote{The definitions of ``fast'' and ``slow components'' are of common use in the community and generally refer to the light emitted from singlet and triplet decay, respectively. However, they are also used, in analysis frameworks, to indicate the two parts of the integrated light pulse mostly dominated by the singlet and triplet dimer populations. Starting with section~\ref{sec:analysis_xArapuca}, the definitions of fast and slow component within our analysis framework will be given and highlighted with italicized text.}).

\begin{figure}[ht]
    \centering 
    \includegraphics[width=.8\textwidth,]{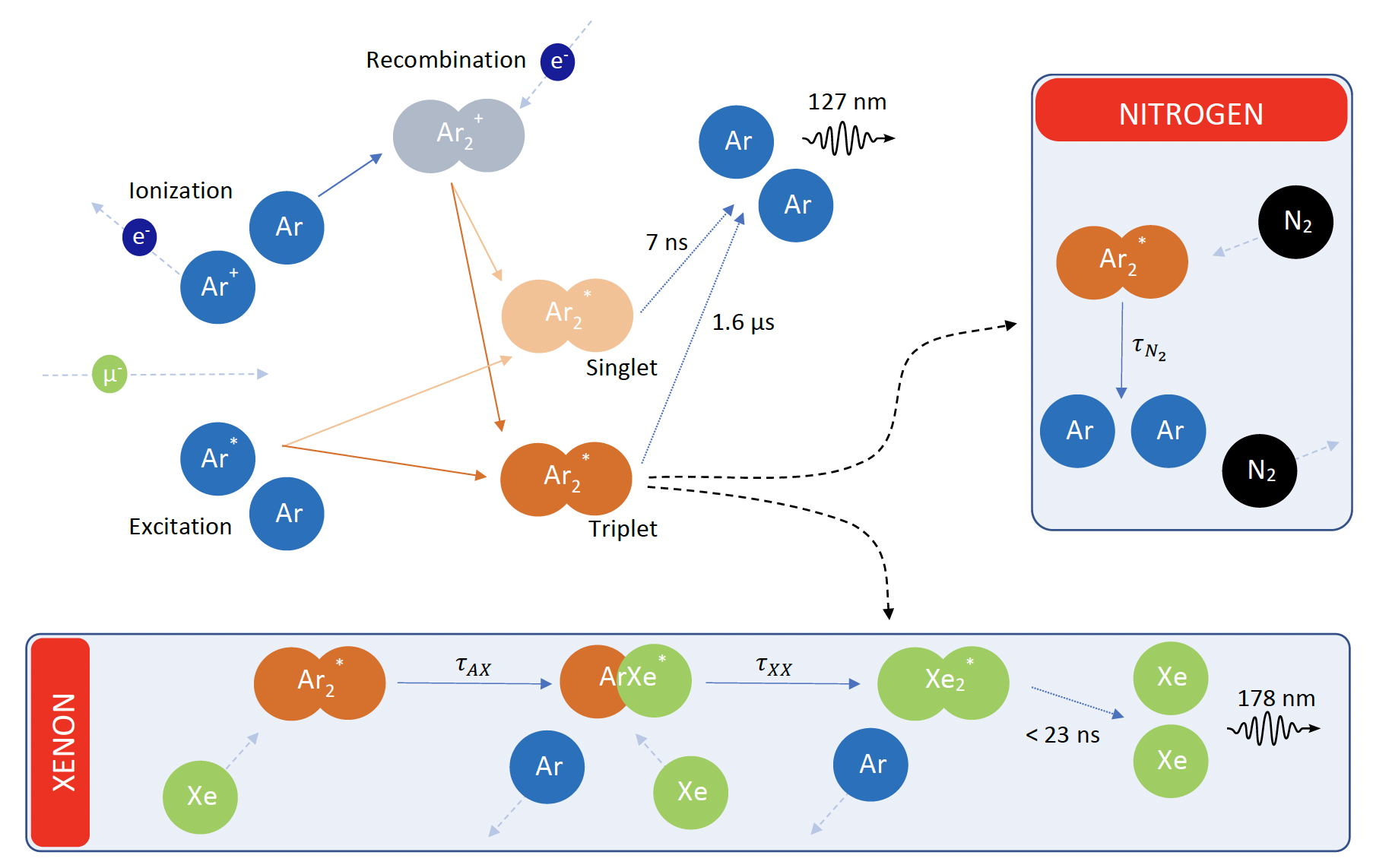}
    \caption{\label{fig:xe-transition} Schematic representation of the production process of scintillation light in pure liquid argon, and the way it is affected by xenon doping and nitrogen quenching. The time constants of the non-radiative energy-transfer processes $\tau_{AX}$ and $\tau_{XX}$ depend on the xenon concentration in LAr.}
\end{figure}

As shown in figure~\ref{fig:xe-transition}, in the presence of xenon, a non-radiative collision of a first xenon atom with the dimer leads to the formation of a new hybrid dimer ArXe$^*$, whereas the interaction of a second Xe atom yields a full transfer of energy to a Xe$_2^*$ dimer, which is at this point the entity decaying with emission of light at 178~nm. The time constants of these two transition processes, identified in figure~\ref{fig:xe-transition}, are defined as follows: $\tau_{AX} \times [Xe] \sim \SI{5.3}{\micro\second \times \ppm}$ and $\tau_{XX} \times [Xe] \sim \SI{20}{\micro\second\times\ppm}$ \cite{Buzulutskov_2017}, and they depend directly on the xenon concentration. At relatively low concentrations, below 1~ppm, the double interaction has a low enough probability to let a certain number of hybrid dimers ArXe$^*$ survive long enough to de-excite, producing an intermediate light component around 150~nm, as shown in~\cite{Neumeier_2015_150nm}. This hybrid component is expected to disappear as the concentration increases to a few~ppm~\cite{Neumeier_2015_150nm}.

In the presence of xenon the number of photons emitted from the long-lived triplet state of the Ar$_2^*$ dimer ($^3\Sigma^+_u$) drops significantly, as the dimer is destroyed by the collision with xenon atoms, before decaying. Overall, the total light emitted is characterised by smaller decay-time constants. The characteristic time profile of the scintillation pulse is modified by the presence of xenon, in a way that is proportional to its concentration. This effect will be illustrated in more detail when discussing the data collected in ProtoDUNE-SP in sections~\ref{sec:analysis_xArapuca} and~\ref{sec:analysis_protodune_pds}. 

For ProtoDUNE-SP, a concentration of a few ppm translates into injecting a few kilograms of xenon in the LAr bulk. Therefore, the detector would be a feasible and effective test-bed to study effects of xenon doping and long term behavior of xenon in LAr at a large scale never attempted before. 

\subsection{Effect of nitrogen contamination in LAr}

As discussed in ref.~\cite{Acciarri_2010_N2}, the presence of nitrogen in liquid argon affects scintillation light emission. This is a well-known process called \textit{quenching}, where the non-radiative collisional reaction $Ar_2^* + N_2 \rightarrow 2Ar + N_2$ destroys the argon triplet excimers before de-excitation. On the other hand, nitrogen does not affect light transport as it is transparent to wavelengths above 100~nm~\cite{Acciarri_2010_N2}. 

Due to the nature of the xenon interaction with the long-lived triplet state argon dimers, this process would be expected to be competitive with the nitrogen quenching effect~\cite{Buzulutskov_2017}. As a matter of fact, it appears to have a larger interaction cross-section. For this reason, in addition to the beneficial effects already discussed, xenon doping can also help to negate the effects of impurities in liquid argon, recovering light that would otherwise be lost. This was the case for ProtoDUNE-SP, which experienced an unexpected event with an argon recirculation pump that allowed atmosphere (nitrogen) in the liquid argon bulk at a level that significantly affected the photon yield (see section~\ref{sec:n2}).

In the ternary mixture Ar-N$_2$-Xe, the two energy transfer processes are thus in competition for their effects on light production from the decay of the triplet Ar$^{\ast}_2$ dimer (see figure~\ref{fig:xe-transition}): non-radiative interactions with a quencher, like N$_2$, effectively suppress light production through dimer destruction. On the other hand, xenon interactions simply shift the excitation energy to different molecules (ArXe$^*$ first, Xe$^{\ast}_2$ later). These are usually able to decay even in the presence of nitrogen, thanks to shorter decay constants, with respect to the argon triplet. Overall, the light output from the mixture is strongly dependent on the concentration of both the quencher and the dopant.
A more detailed discussion of the modeling of the ternary mixture and its characterization in large volume LArTPCs is deferred to a later publication.

\section{The ProtoDUNE Single-Phase detector}
\label{sec:ProtoDUNE-SP}
The ProtoDUNE single-phase LArTPC is a prototype for the first module of DUNE~\cite{Abi_2020_V4}, exploiting full-scale detector elements. With a total LAr mass of up to 770~t (the actual mass during the xenon doping run was of 720~t), it is the largest single-phase LArTPC detector built to date. It is located in the dedicated extension of the EHN1 hall in CERN North Area, where a tertiary portion was added to the existing H4 beam-line, to provide very low-energy charged-particle beams, as part of the CERN Neutrino Platform program. Construction, installation, and commissioning of the ProtoDUNE-SP detector was completed in July 2018, and is reported in ref.~\cite{pdsp_design}. Immediately after LAr filling and detector activation, beam data were collected in the 0.3-7~GeV range from September to November 2018~\cite{Abi_2020_PDSPperf}. After the beam run, it operated until July 2020 collecting data with cosmics, to validate the design solutions for the future DUNE far detector modules, demonstrate operational stability, and eventually to perform R\&D on different aspects of LArTPC technology. Doping LAr with xenon to enhance the light collection of the photon detectors, as presented in this paper, was part of these R\&D efforts. 

\begin{figure}[ht]
    \centering 
    \includegraphics[width=.7\textwidth,]{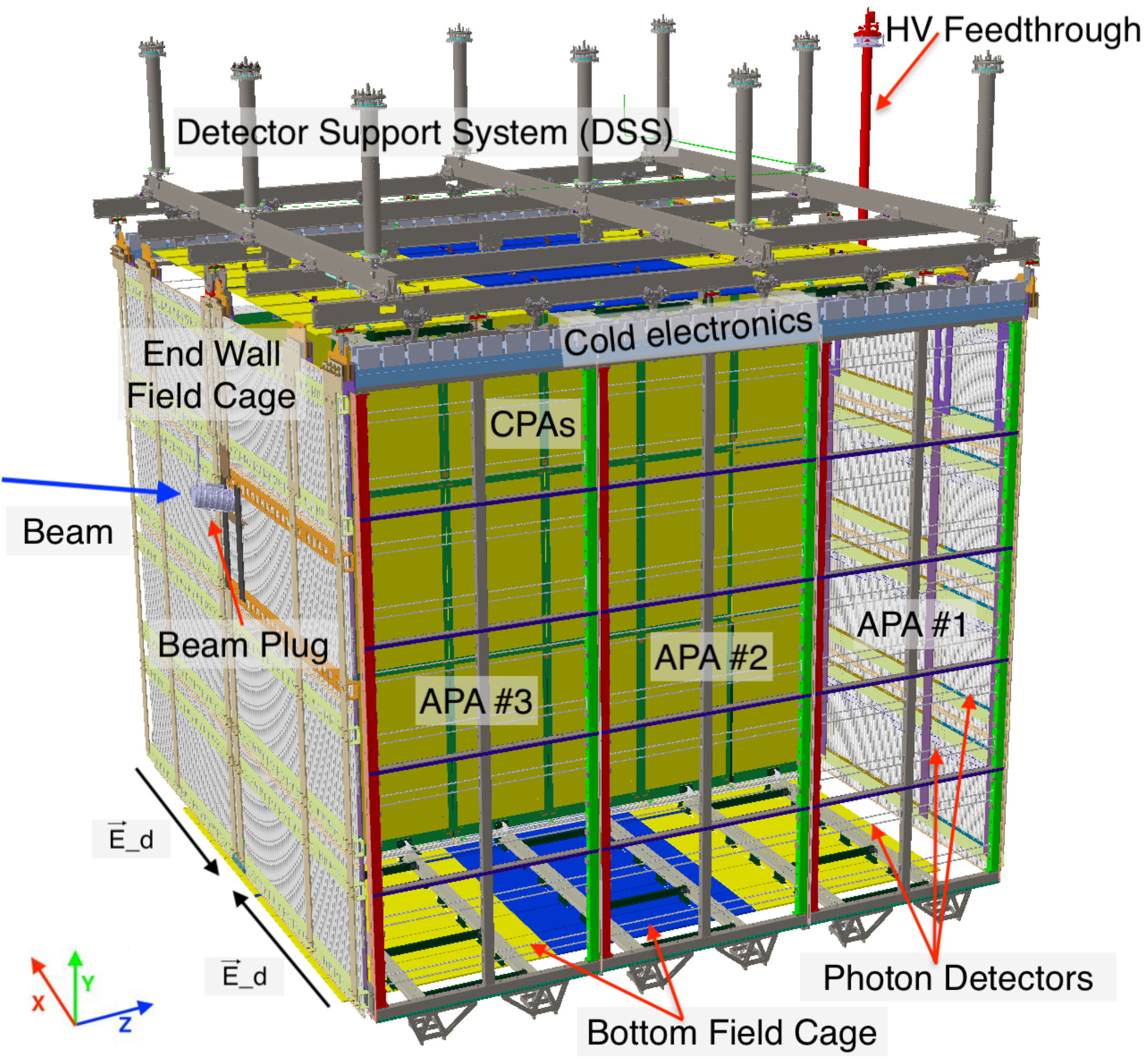}    
    \caption{\label{fig:PDSP_detector} 3D model of the ProtoDUNE-SP detector with labelling of all major components and definition of coordinate system used (bottom left).}
\end{figure}

The ProtoDUNE-SP TPC is described in all its components in ref.~\cite{pdsp_design}. It features 410~t of active LAr volume with dimensions of 7.2~m $\times$ 6.0~m $\times$ 6.9~m. As shown in figure~\ref{fig:PDSP_detector}, the active volume is split in two by a central cathode plane made of three \textit{cathode plane assemblies} (CPAs), defining two identical volumes, each with 3.6~m of drift length. The cathode is biased to -180~kV, providing a nominal 500~V/cm electric field in the drift region. On both sides of the cathode, at a distance of 3.6~m, the \textit{anode planes assemblies} (APAs) are installed. Each APA is made up of four layers of wire planes: three active planes for charge readout, plus a grounded ``grid'' wire plane in front of them. Each drift volume is read-out by three APAs.
The two volumes are called Left chamber and Right chamber, according to their position along the direction of the incoming charged-particle beam. 
 
\subsection{Photon detection system}
\label{sec:PDS}
\begin{figure}[ht]
    \centering 
    \includegraphics[width=.6\textwidth,]{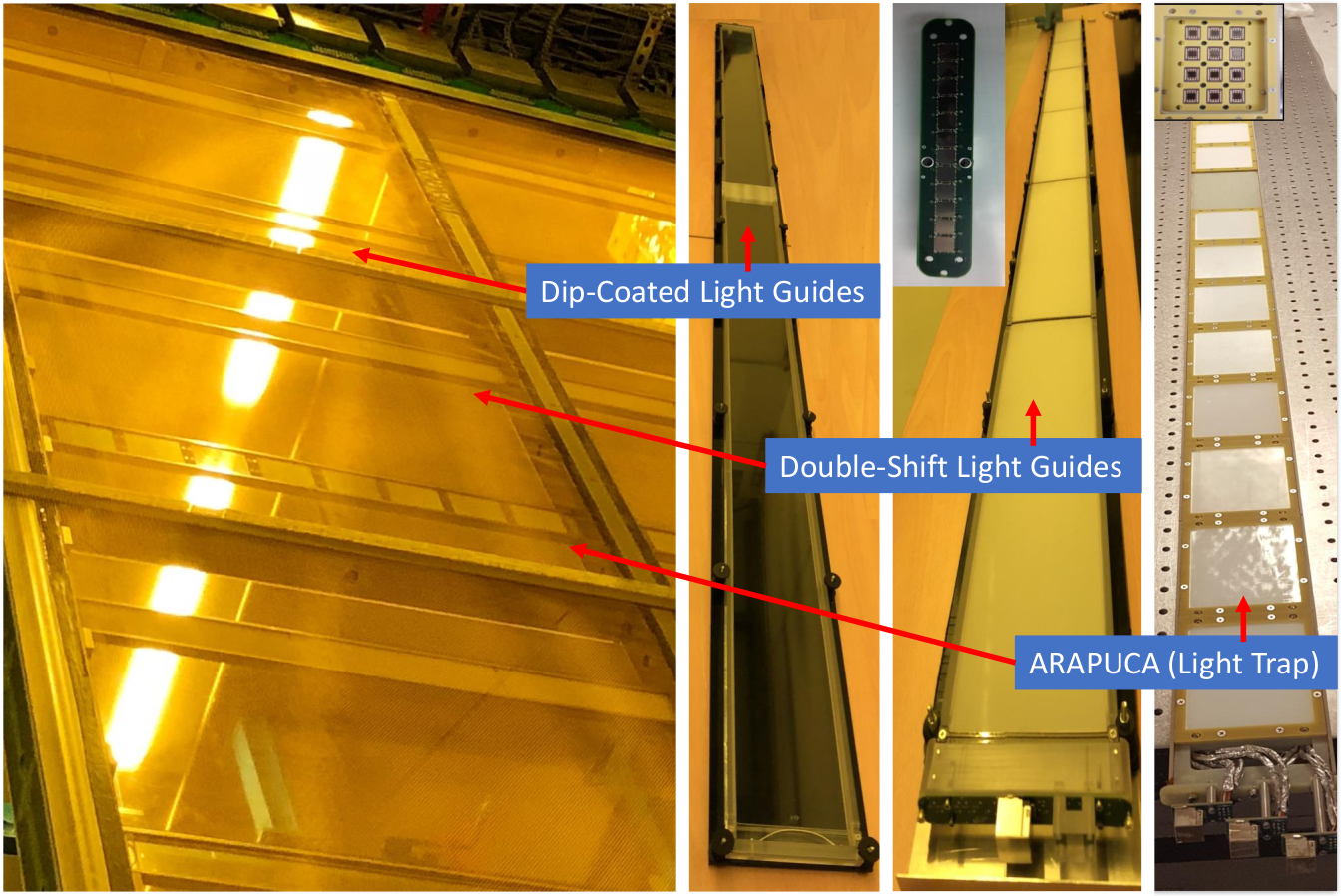}
    \caption{\label{fig:PDSP_PDS}The three technologies of PDS modules shown inside the APA frame and individually for comparison.}
\end{figure}

The scintillation light produced by charged particles traversing the LAr is recorded by the photon detection system (PDS), which is made of 60 optical modules of active area $207 \times 8.6$~cm$^2$ each. Ten modules are inserted into each APA frame, facing the TPC drift volume and regularly spaced along the vertical direction. Each module combines a photon collector and a photon sensor. Three different collector designs were implemented in ProtoDUNE-SP: ``double-shift light guides'' (DSLG)~\cite{Howard_2017dqb}, ``dip-coated light guides'' (DCLG)~\cite{LBugel,ZMoss}, and ARAPUCA light traps~\cite{Machado_2016}. Silicon photomultiplier arrays from Hamamatsu and SensL vendors are deployed as sensors: models and versions of such sensors are described in detail in ref.~\cite{pdsp_design}. Locations of the PDS modules in an APA frame and the three types of detector technologies are shown in figure~\ref{fig:PDSP_PDS}, whereas their performance is illustrated in detail in ref.~\cite{pdsp_design, Abi_2020_PDSPperf}.

\subsection{Cosmic-Ray Tagger}
The ProtoDUNE-SP detector is exposed to a flux of $\sim$180~cosmic muons/(m$^2$~s). A fraction of these particles is tagged by a cosmic-ray tagger (CRT,~\cite{pdsp_design}): this is made of scintillator counters (strips) read by SiPMs, and it consists of four large assemblies, two mounted upstream and two downstream of the cryostat. Each assembly covers an area approximately 6.8~m high and 3.65~m wide. Modules are instrumented with 64 scintillator strips 5 cm wide and 365 cm long. Two-dimensional sensitivity is achieved by putting together groups of four modules into assemblies, with two modules being rotated by 90$^{\circ}$ with respect to the other two. 
A CRT track is reconstructed by drawing a line from hits in strips of the upstream modules to hits in strips of the downstream modules, the muon time-of-flight information dictating the width of the relative coincidence window \cite{CRT_diurba, Abi_2020_PDSPperf}. Coincidences between upstream and downstream CRT modules are used in the trigger configuration of ProtoDUNE. By changing such configuration, i.e. selecting different CRT modules for the coincidence, it is possible to select different sets of cosmic-ray muons, with a well defined direction (e.g., parallel to APAs), time stamp, and average distance from the anode plane.

\subsection{The X-ARAPUCA detectors in ProtoDUNE-SP}
\label{sec:xArapuca_telescope}

The ARAPUCA technology is based on light trapping, as discussed in ref.~\cite{Machado_2016}. In the base concept, trapping of UV photons is achieved as follows: 127~nm photons hitting the detector are shifted to 350~nm by a thin p-terphenyl (pTp) coating located on top of a dichroic filter, that features a 400~nm transparency cutoff. A second coating layer, with TPB, converts 350~nm photons to 420~nm. 
The upgrade of the technology (X-ARAPUCA~\cite{Machado_2018}) replaces the second coating layer with a WLS light guide, enhancing detection efficiency\footnote{For the photon detection system modules addressed in this paper, unless otherwise noted,  ``detection efficiency'' is defined as the ratio of the number of photon detected by the photosensor to the number of photons impinging on the sensitive surface of the module.}. In both versions, the produced 420~nm photons are trapped inside the detector by the filter, fully reflective above the 400~nm cutoff, and bounce back-and-forth until they reach the photosensors (cryogenic SiPMs).

Two ARAPUCA modules were installed in ProtoDUNE-SP for the first beam run; performance studies report a measured detection efficiency of 1-2\%, as defined in ref.~\cite{Abi_2020_PDSPperf}. 
Two early prototypes of an upgraded version of these detectors, called X-ARAPUCAs, were deployed in ProtoDUNE-SP expressly for the xenon doping run.

\begin{figure}[ht]
    \centering
    \includegraphics[width=0.65\textwidth,keepaspectratio]{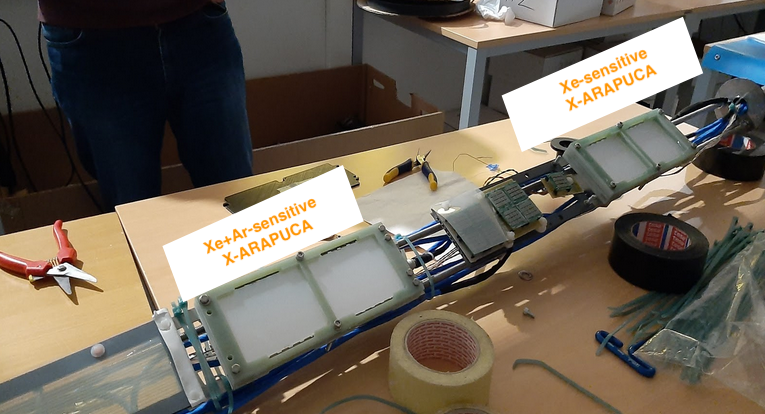}
    \caption{X-ARAPUCA detectors installed on a dedicated support and ready for insertion in the ProtoDUNE-SP cryostat.}
    \label{fig:new_pd_setup}
\end{figure}

\begin{figure}[htbp]
    \centering
    \includegraphics[width=0.54\textwidth,keepaspectratio]{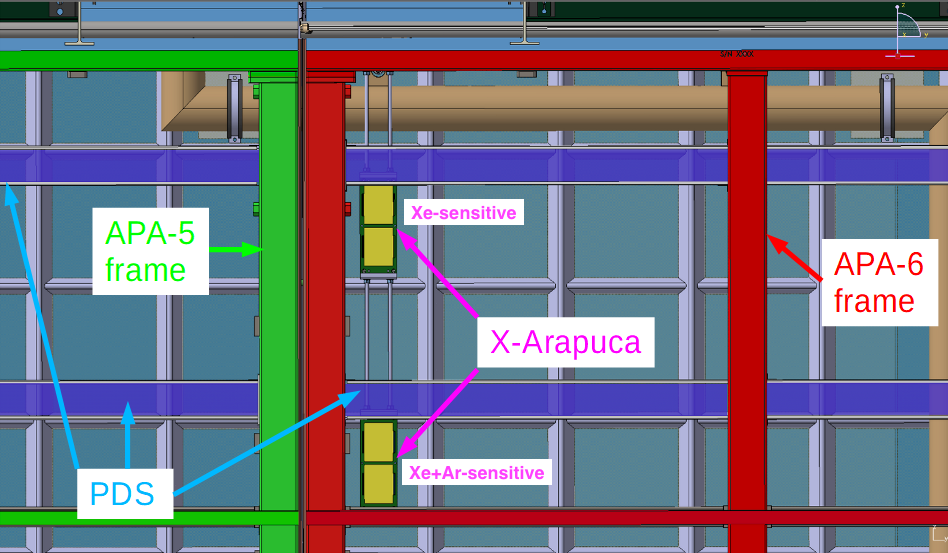}
    \includegraphics[width=0.36\textwidth,keepaspectratio]{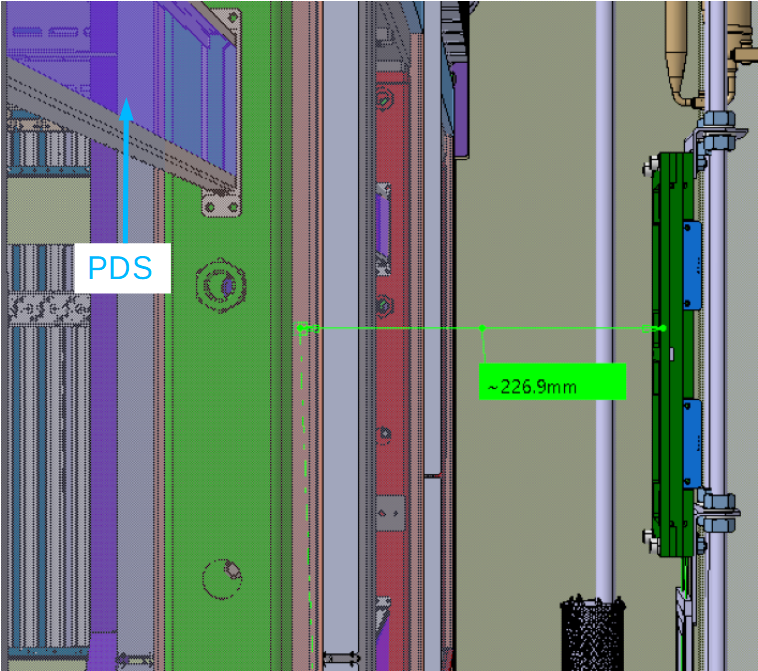}
    \caption{3D model of the two X-ARAPUCA detectors inside the ProtoDUNE-SP cryostat. Left: front view. In green, the frame of APA-5, in red the frame of APA-6, in blue the PDS bars. Right: side view, showing the position of the X-ARAPUCAs with respect to the APA frames and PDS.}
    \label{fig:telescope_cad}
\end{figure}

The two X-ARAPUCA (XA) detector units were installed on a dedicated support (see figure~\ref{fig:new_pd_setup}). They are placed behind APA-6, upstream with respect to the beam, at a distance of 22.7~cm from the anode frame (see figure~\ref{fig:telescope_cad}).
The trigger for these detectors is not connected to the main ProtoDUNE DAQ. Instead, it is obtained from cosmic rays, through a standard triple coincidence of $15.5 \times 44$~cm plastic scintillators, located on the cryostat roof, 1.15~m far from the active volume.
The three paddles select a solid angle of $\sim$0.43~steradians, resulting in an average trigger rate of about 1~Hz.

The two detectors are identical but for the addition, on the top one, of a fused silica window, which is opaque to 127~nm radiation, whereas it has a measured transparency of $\sim$80\% for 178~nm photons~\cite{souza2021arapuca}. For this reason, this unit collects only light from xenon de-excitation and will be labeled in the following as ``Xe-XA''. The bottom detector is instead sensitive to both argon and xenon light, and it will be referred to as ``Ar+Xe-XA''.

The X-ARAPUCA light detection efficiency in liquid argon was first measured in two prototypes, one $10\times 8$~cm$^2$ in size at Unicamp, Brazil~\cite{Segreto_2020} and the other $20\times 7.5$~cm$^2$ in size at INFN Milano-Bicocca, Italy~\cite{Brizzolari:2021akq}: the latter is of the same type and size of those deployed in this work. From these tests, an average detection efficiency of~$\sim$2.3\% is obtained.

Both X-ARAPUCAs installed on ProtoDUNE-SP are equipped with Hamamatsu MPPCs S13360-6050VE~\cite{SiPM_spec} with a $6\times6$~mm$^2$ active area and 1.3~nF terminal capacitance.
They were operated with a bias of 47.8~V, or +4.8~V over-voltage (VoV, i.e. above the SiPM breakdown voltage). This value was chosen to guarantee the SiPMs PDE >50\% and to partially compensate for the lack of a cryogenic front-end amplifier.
Each detector features two windows, both equipped with two arrays of four SiPMs positioned against the long sides of the WLS bar: the SiPMs within each array are readout in parallel, resulting in 4 readout channels per detector, read out via CAT6 cables\footnote{During data-taking, only six channels out of eight were operational and thus used for the analysis reported in this paper.}. Readout is performed by a customized version of the standard \textit{SiPM signal processor} (SSP) board, used for the first run of ProtoDUNE-SP~\cite{Abi:2271524}.

\section{Cryogenics operations for xenon doping in ProtoDUNE-SP}
\label{sec:cryostat}

The ProtoDUNE-SP cryostat contains 720~t of LAr at 87.5~K, that is continuously purified through a cooling-recirculation plant. The cryostat and the cryogenic plant are described in detail in ref.~\cite{pdsp_design}.

\begin{figure}[ht]
    \centering 
    \includegraphics[width=.85\textwidth,]{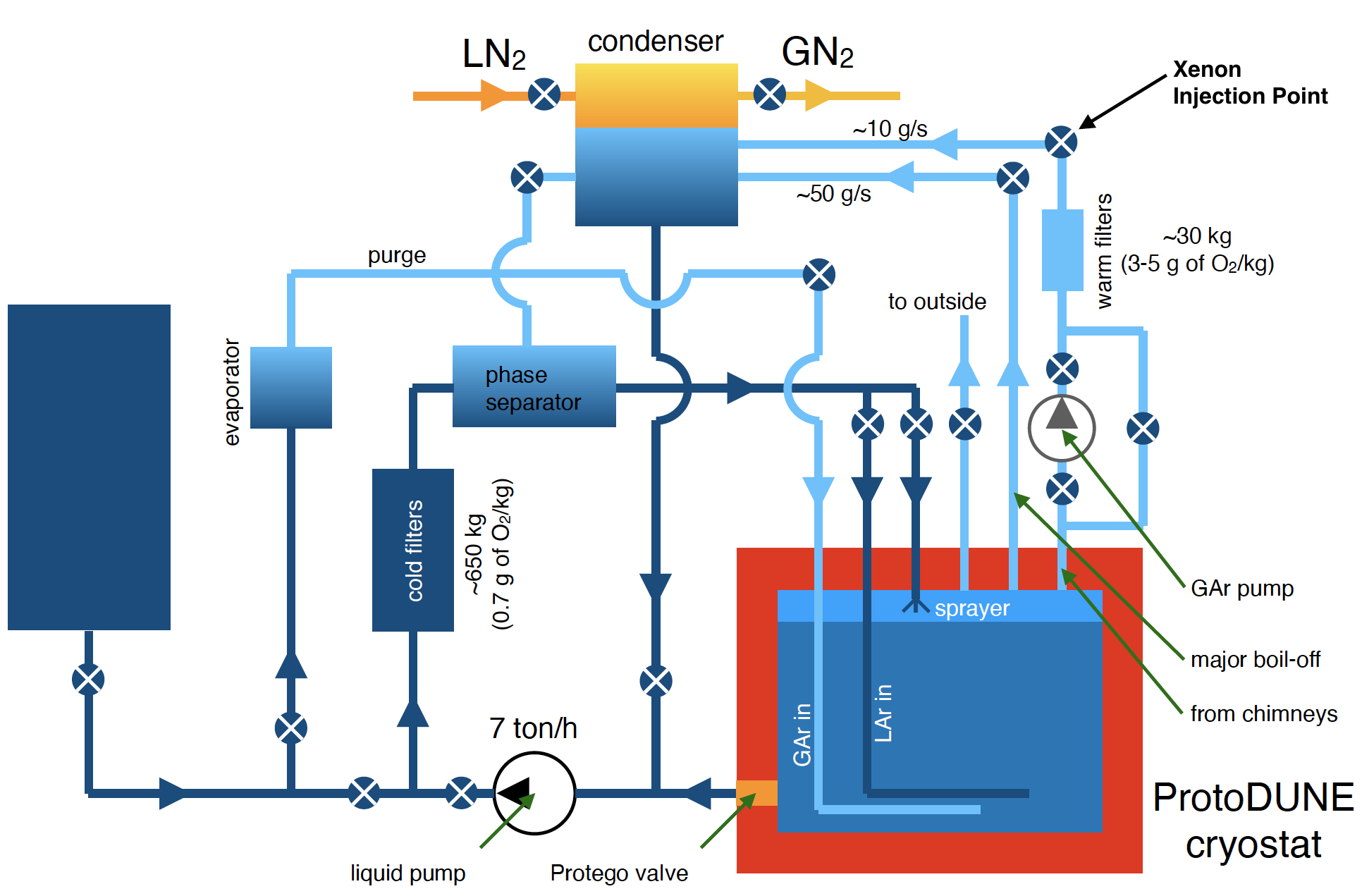} 
    \caption{\label{fig:np04_cryo}Schematics of the ProtoDUNE-SP cryogenic system.}
\end{figure}

The system layout is depicted in Figure~\ref{fig:np04_cryo}. It consists of two main circuits, one for liquid and one for gas recirculation.
The first circuit extracts LAr at the bottom of the cryostat by means of a cryogenic pump. The liquid is then forced through a cold purifier at a rate of $\sim7$~t/hour. The purifier consists of a first section filled with molecular sieve optimized to remove polar molecules, such as H$_2$O or CO$_2$, and a second section containing copper deposited on alumina pellets, which adsorbs O$_2$~\cite{CURIONI2009306}. The purified liquid is injected back at the bottom of the cryostat at a slightly warmer temperature (a fraction of degree) that allows upward diffusion, thus ensuring a better mixing with the bulk LAr in the cryostat.

The gas circuit employs the same filter cartridges described for the liquid circuit. It is meant to both stabilize the operating pressure in the cryostat, by re-condensing the boil-off gas continuously produced by the residual heat input, and to purify the argon gas present in the ullage and in the feed-through chimneys. Indeed, these areas are expected to be heavily polluted, due to the degassing of materials (mainly the cables) present in this area. The re-condensed gas is then mixed with the liquid extracted from the LAr bulk. 

As xenon solidifies at 161~K, the creation of a solution with liquid argon can be obtained only with extreme care, in order to avoid its freezing.
Preliminary tests performed by the collaboration at CERN, with smaller LArTPC prototypes equipped with gas recirculation/purification systems, demonstrated that xenon can be efficiently mixed with argon by injecting it in the gas phase, before the recondensation.
Several mixing ratios were tested, showing that the Ar-to-Xe mass ratio must be above $10^3$ to avoid solidification of the xenon on the walls of the condenser. This \textit{freeze-out} effect is observed since, at the highest xenon concentrations, the pipes of the condenser get clogged up and the argon recirculation stops.

\subsection{Nitrogen contamination}
\label{sec:n2}

\begin{figure}[ht]
    \centering
   \includegraphics[width=.65\textwidth,trim=0 0 0 0,clip]{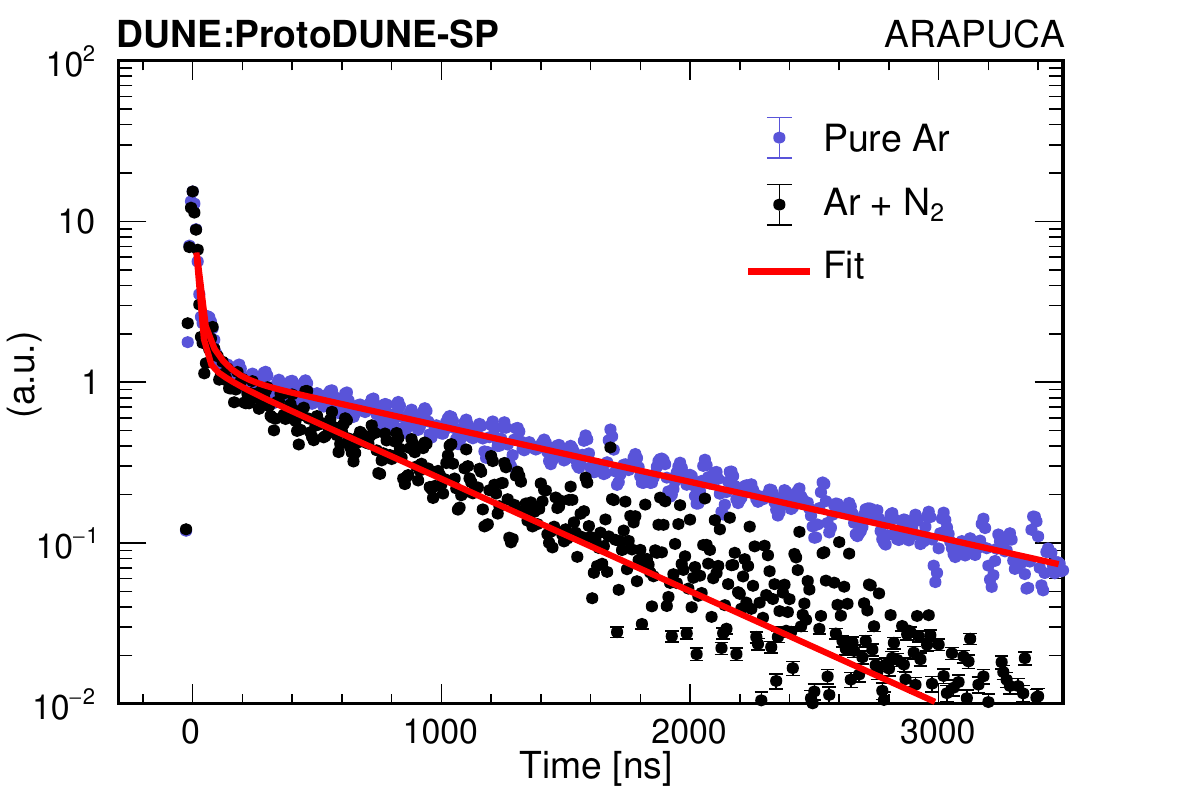}
    \caption{\label{fig:02_PureArVSArN2}Typical scintillation light waveforms from a ProtoDUNE-SP ARAPUCA module.
    Blue: ``pure'' Ar (before the air contamination), Black: after air contamination and purification (only N$_2$ contaminant is present). The pure argon waveform is scaled to have the same maximum amplitude on both pulses in the fast component region.}
\end{figure}

During the long cosmic run of ProtoDUNE-SP, a sudden failure in the gas recirculation pump occurred, injecting a non-negligible amount of air inside the detector. Molecules like O$_2$, CO$_2$, and H$_2$O were efficiently removed by the purification system, during the following three weeks of recirculation through the filters. However, the system cannot remove N$_2$, which remained in the detector until the end of the run. 
As mentioned in section~\ref{sec:xe_doping}, nitrogen suppresses scintillation light emission, through the process of quenching.
This effect is demonstrated in figure~\ref{fig:02_PureArVSArN2}, which shows the typical profile of the scintillation light pulses for non-polluted LAr and LAr $+$ N$_2$ after contamination, as obtained from ProtoDUNE-SP data (specifically from the ARAPUCA module installed in APA 6).

By measuring the value of the decay time constant of the argon triplet scintillation light component in both conditions~\cite{Heindl_2011}, we can compute~\cite{Acciarri_2010_N2} the total amount of N$_2$ that is present in LAr:~5.4$\pm$0.1~ppm, and derive the quantity leaked in during the accident: 5.2$\pm$0.1~ppm. The initial (pre-accident) concentration estimated with this method is $\sim$0.2~ppm N$_2$: this is compatible with the data provided by the LAr supplier (AirLiquide\footnote{https://www.airliquide.com/}) and with the values obtained from direct measurements performed during argon deliveries.

\subsection{Xenon doping campaign of ProtoDUNE SP}
\label{sec:xe_doping_schedule}

The xenon doping campaign of ProtoDUNE-SP started in February 2020 and lasted five months, with the goals of: $(i)$ studying light emission in the presence of xenon; $(ii)$ investigating long term stability and uniformity of the doped xenon inside the cryostat, and $(iii)$ checking for possible effects of xenon on TPC charge response. This campaign became even more important after the unexpected nitrogen pollution event described above, given the competing effect of xenon with respect to the nitrogen quenching (see section~\ref{sec:xe_doping}). 

The xenon injection point is placed along the chimney boil-off recirculation line (see figure~\ref{fig:np04_cryo}), after the gas purification filter but some distance before the condenser, to allow for full mixing within the gas flow. The maximum xenon injection mass flow rate was set to 36~g/h, to be well within the Ar-to-Xe mass ratio limit described above; this corresponds to 50~ppb/hour in the ProtoDUNE-SP detector. Based on a numerical (CFD\footnote{Computational Fluid Dynamics}) simulation of the LAr flow within the ProtoDUNE cryostat, the xenon injected at this rate is expected to be uniformly distributed in LAr within few hours.
A detailed description of all steps of the doping procedure, and the lessons learned while performing it, is reported in appendix~\ref{sec:appendix_inj}.
 
The run consisted of six injections, however the last two were performed consecutively over a few days so they are treated as one doping step in the analysis. The amount of xenon injected in each step and the corresponding concentration inside the cryostat is summarized in table~\ref{tab:Xe_ppms}.
Combining all the doping steps, we injected 13.6~kg of xenon into the cryostat. In the 720~t of LAr in ProtoDUNE-SP, this translates to 18.8~ppm of xenon concentration by mass, the conversion factor being 1~ppm $\simeq$ 720~g of xenon.

\begin{table}[htbp]
\centering
\caption{\label{tab:Xe_ppms} Six xenon doping steps in ProtoDUNE-SP. The dates, injected xenon mass in grams, and concentration in ppm by mass are given for each doping step.}
\smallskip
\begin{tabular}{|c|c|c|c|}
\hline
$\bf{Injection \ Number (\#)}$ & \bf{Date} &$\bf{Injected \ Xe[g]}$ &$\bf{Injected \ Xe[ppm]}$ \\\hline
1   & 13-14 February 2020  & 776  &1.1 \\
2   & 26-28 February 2020  & 2234  &3.1 \\
3   & 3-8 April 2020  & 5335  &7.4 \\
4   & 27-30 April 2020  & 3192  &4.5 \\
5   & 15-16 May 2020  & 400  &0.6 \\
6   & 18-20 May 2020  & 1584  &2.2 \\
\hline
\end{tabular}
\end{table}

Extensive data taking was performed during each injection and between the dopings, both with the ProtoDUNE photon detection system and with the supplemental X-ARAPUCA modules.
The evolution of the scintillation light emission was monitored during the whole campaign, as a function of the amount of injected xenon.

\section{Analysis of the X-ARAPUCA data}
\label{sec:analysis_xArapuca}

In this section we introduce the data collected with the dedicated X-ARAPUCA detectors, which cover the period from the first xenon injection onward.

The X-ARAPUCA data are acquired with a standalone SSP board that communicates with a local DAQ system that collects and saves data. When a cosmic ray trigger occurs (see section~\ref{sec:xArapuca_telescope}), the SSP starts digitizing the input signals coming from the SiPMs. The board implements a digitizer that samples at \SI{150}{\mega\hertz} with a 14 bit resolution and an aggregator that streams out 2000 samples waveform for each trigger. The sampling time, defined as a ``tick'', corresponds to 6.67~ns, roughly translating into a $\SI{13.3}{\micro\second}$-long acquisition window. To have a proper baseline estimation, the pre-trigger is chosen to be 240 ticks, i.e. \SI{1.6}{\micro\second}.

At the beginning of the first run, an unexpected source of noise was found to be generated by the trigger electronics. In order to mitigate this noise, a subset of triggered events with no detectable physical signal was identified and their recorded pulses were averaged. These empty events are due to crossing muons that trigger the system, but interact early in the cryostat roof, therefore producing no detectable light in liquid argon. We employed such averaged empty triggers to subtract this noise feature from the candidate signal waveforms. We monitored the effect during the runs and verified that it remained stable throughout the whole data-taking campaign.

\begin{figure}[ht]

    \centering 
            \includegraphics[width=1.\textwidth,trim=0 0 0 0,clip]{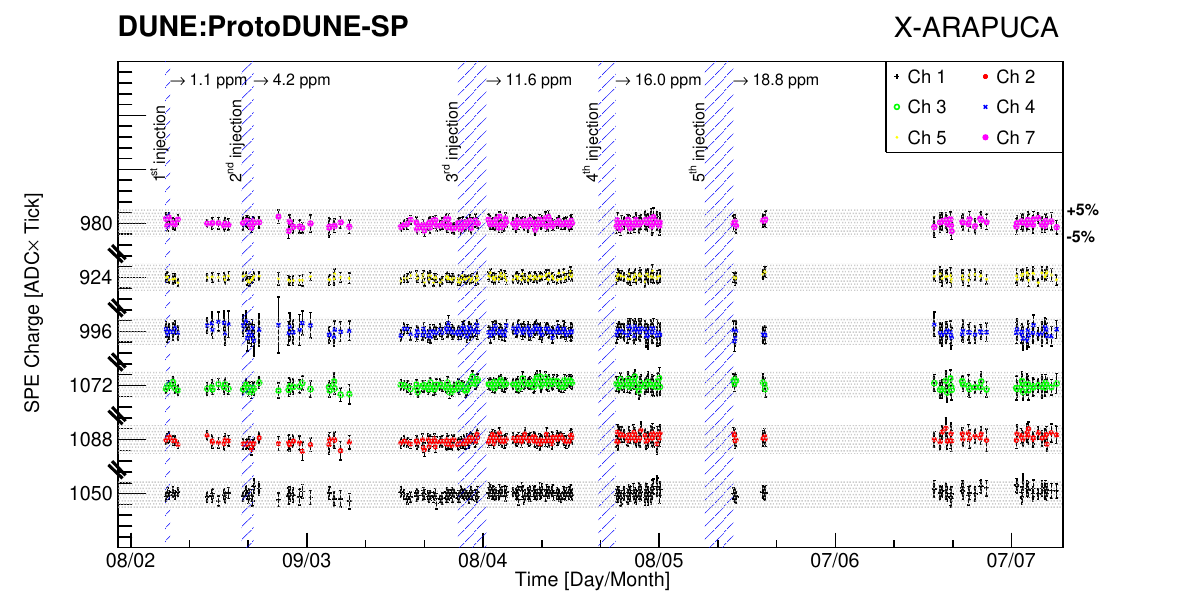}
    \caption{\label{fig:spe_stab} Mean SPE charge stability for all runs and each channel. Six out of eight X-ARAPUCA channels are shown, since one channel per detector was not functioning properly and was excluded. Channels 1,2,3 refer to the ``Ar+Xe'' X-ARAPUCA, whereas channels 4,5,7 pertain to the ``Xe'' X-ARAPUCA. Values on the y-axis show the SPE charge average across all the data for each channel. Gray bands highlight $\pm5\%$ relative variations with respect to the average value. Runs cover an overall six-month doping period, colored areas represent specific dopings.}
\end{figure}

Monitoring of sensors and electronics was carried out during the entire acquisition period, by analyzing the single photoelectron (SPE) response of the system. A peak finder algorithm searches photoelectron pulses in the tail of each acquired signal, i.e. well beyond the triggered pulse. The integral of this sub-sample of data is then histogrammed. The resulting distribution exhibits a first peak that represents the pedestal (events with no photoelectrons), whereas the following $n^{th}$ peak represents respectively $n$ photoelectrons. The first two peaks are fit with two Gaussians and the difference in their mean values is taken as the SPE charge. Figure~\ref{fig:spe_stab} shows its stability along the entire run. The outcome of these quality tests demonstrated that the X-ARAPUCA system ran in stable conditions during the entire doping campaign.

\subsection{Data selection and deconvolution}

The data acquired with the X-ARAPUCA detectors were first converted into a ROOT~\cite{BRUN199781,fons_rademakers_2020_3895855} TTree and pre-processed applying a moving average filter to reduce the white noise and subtracting the baseline. For each waveform the integral, peak amplitude in ADC counts, and peak time are computed and recorded.

Data are selected applying two main quality cuts: first, saturated events are discarded imposing a maximum on the peak-height parameter associated to each waveforms. This threshold value takes into consideration the electronics saturation level. Secondly, events with an ill-defined baseline or with a relevant pileup are removed: these are events where a scintillation signal is present in the pre-trigger region (0 -- 1.33~\si{\micro\second}), or in the final part of the waveform (8.66 -- 13.33~\si{\micro\second}), respectively. The waveforms are discarded if 10~or more photoelectrons are found in the regions defined above. 

The waveforms passing these cuts are averaged to reconstruct the response function of the detector. The information enclosed in these waveforms is the convolution of three main effects, $ S(t) = L(t) \otimes XA(t) \otimes h(t)$: the scintillation light time-profile $L(t)$, the X-ARAPUCA $XA(t)$ time-response and the electronics $h(t)$ response. The first is characterized by the light output, the emission properties of the mixture (Ar+Xe+N$_2$) and by the light propagation including absorption and Rayleigh scattering. The second is characterized by the X-ARAPUCA response, in particular by the absorption and re-emission of the wavelength shifters. As the re-emission delay of TPB and pTP is below $<\SI{10}{\nano\second}$~\cite{FLOURNOY1994349}, we can consider the time dependence of this effect negligible with respect to the other time constants involved. The third effect $h(t)$ is due to the response of both sensors and the electronics to a single photon signal. To retrieve the scintillation signal $L(t)$ containing the relevant physical information, this last effect needs to be deconvolved as the most relevant. In fact, the signal coming from SiPMs is proportional to the number of photons but has a time duration of about $\SI{400}{\nano\second}$, comparable with scintillation signals.

To deconvolve this effect, a (time-dependent) template for the single photoelectron is needed. A filter for peak finding is implemented to search for single photoelectrons in the pre-trigger region. Once selected, they are aligned at the same time and averaged; the resulting shape is then fitted. The fit function consists of a double exponential convoluted with a Gaussian to account for white noise: $h(t) = Gaus(t; \mu, \sigma) \otimes \left( \exp{[-t/\tau_1]} - \exp{[-t/\tau_2]}\right)$. The two time constants represent respectively the SiPM avalanche discharge ($\tau_1\sim \SI{400}{\nano\second}$) and the electronics shaping time.

More than one deconvolution technique was applied independently on the waveforms, to cross-check the results. One such technique makes use of a custom finite impulse response (FIR) filter to simultaneously de-noise the waveforms and filter out the shape of the single photoelectron response function. The filter\footnote{For the interested reader: with reference to the cited paper, this filter lacks a zero-area requirement. It is a finite-length cusp-like filter with a 33~ns flat top and the cusp shape parameter $\tau _s$=33~ns.} employed is analogous to the one presented in ref.~\cite{FIR}. This algorithm was the one adopted for subsequent analyses: it was tailored for each of the six operating channels to properly take into account the individual exponential decay of channel response function.
Another technique is based on the Gold algorithm~\cite{MORHAC1997113} and the parameters were tuned to optimize the reconstructed singlet component of LAr, while at the same time minimizing the noise.

\subsection{Effects of xenon on LAr light}

Before addressing the analysis results, it should be noted that, in this and the next section, we will often refer to the \textit{fast} and \textit{slow} components of the discussed light pulses. Within our analysis framework, they are defined as follows: the \textit{fast} component is the fraction of the integral of the waveform within the first $\sim74$~ns ($11\times \SI{6.67}{\ns}$~time ticks) after the trigger time. The \textit{slow} component is instead defined as the fraction of the waveform integral starting 11~time ticks after the trigger time. Later on, these definitions will be further discussed.

\begin{figure}[ht]
\centering
    \includegraphics[width=1.0\textwidth,keepaspectratio]{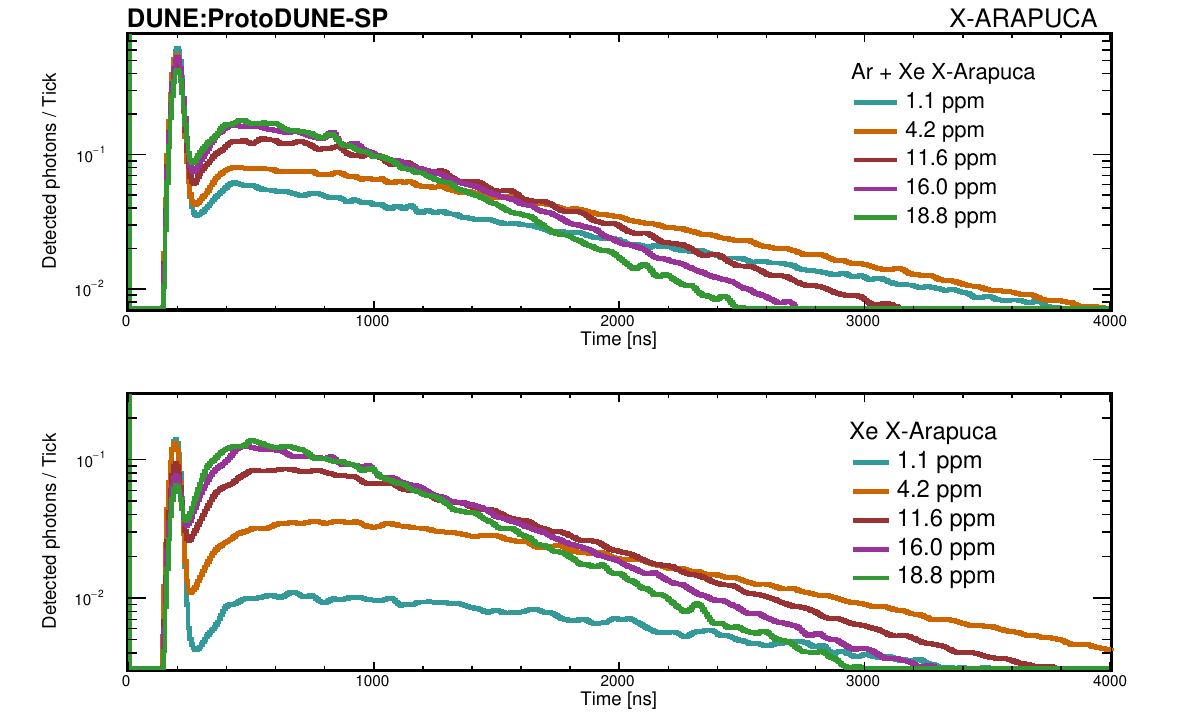}    
    \caption{Average waveforms obtained after deconvolution of single photoelectron pulse, at different stages of xenon doping (after nitrogen pollution). Data from runs with no electric field. Top panel: The Ar+Xe-light sensitive X-ARAPUCA; bottom panel: the Xe-light-only sensitive X-ARAPUCA. Only events with at least three detected photons in the Ar+Xe X-ARAPUCA module are selected.}
    \label{fig:wf_comp}
\end{figure}

\begin{figure}[ht]
\centering
    \includegraphics[width=0.99\textwidth,keepaspectratio]{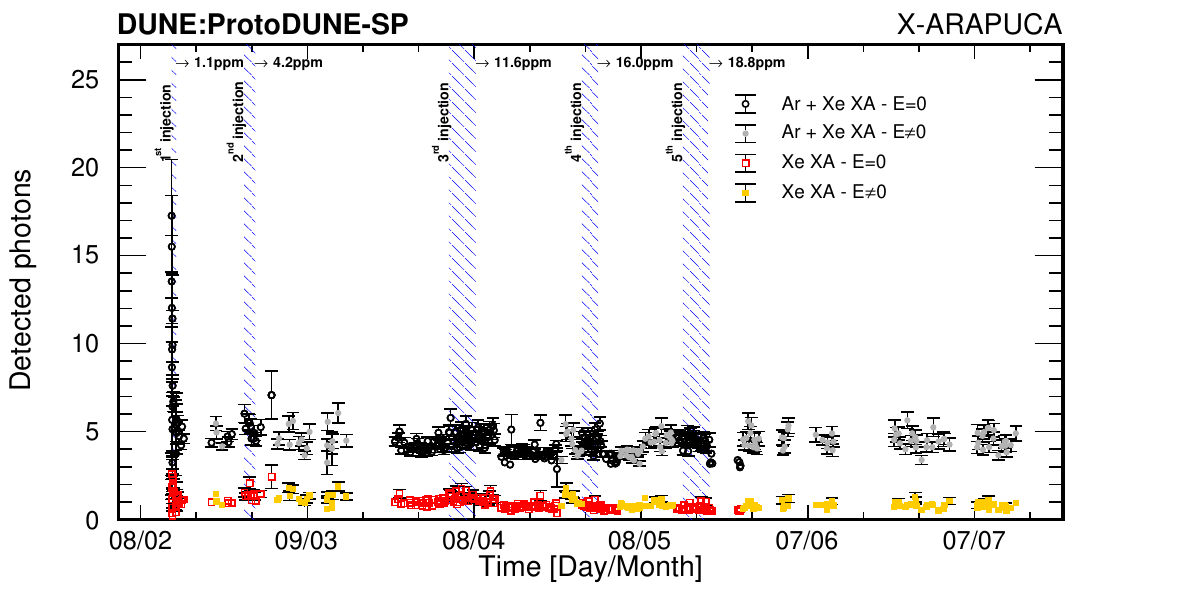}    
    \caption{Time survey of the mean number of photons in the \textit{fast} light component (detected photons with $t<74$~ns after trigger) detected by the Ar+Xe X-ARAPUCA and by the Xe X-ARAPUCA, for runs with and without electric field. Shaded areas indicate xenon injections.
    Only events with at least three detected photons in the Ar+Xe X-ARAPUCA module are selected. }
    \label{fig:LY_fast}
\end{figure}

The effect of the argon-xenon energy transfer, described in section~\ref{sec:xe_doping}, is clearly illustrated in figure~\ref{fig:wf_comp}. The plot shows two sets of light pulses (after deconvolution with the technique introduced in the previous paragraph), one for each X-ARAPUCA, for different xenon concentrations. The overall increase in the amount of light collected (area under the pulse) with increasing xenon concentration is evident in both detectors, as well as the narrowing of the pulse profile. This is ascribed to the transfer of excitation from the argon triplet dimer to xenon dimers. The energy is transferred to xenon before it can be quenched in interactions with nitrogen. As the de-excitation time constant is dominated by the ArXe$^*$ creation process, it is expected to become shorter at higher xenon concentrations when the transfer process is more effective.
These data are collected in the presence of nitrogen where the long tail of the typical argon signal is expected to be strongly reduced (even at the smallest xenon concentrations) with respect to the non-polluted argon case (see figure~\ref{fig:02_PureArVSArN2}). The difference in amplitude between the \textit{fast} component peaks of the two detectors (at all xenon concentrations) is related to the fact that one device is sensitive to the total light (top panel), while the other is only sensitive to 178~nm photons (bottom panel). The evolution of the \textit{fast} peak amplitude with xenon concentration is instead discussed towards the end of this section and shown in figure~\ref{fig:LY_fast}.

\begin{figure}[ht]
\centering
        \includegraphics[width=0.99\textwidth,keepaspectratio]{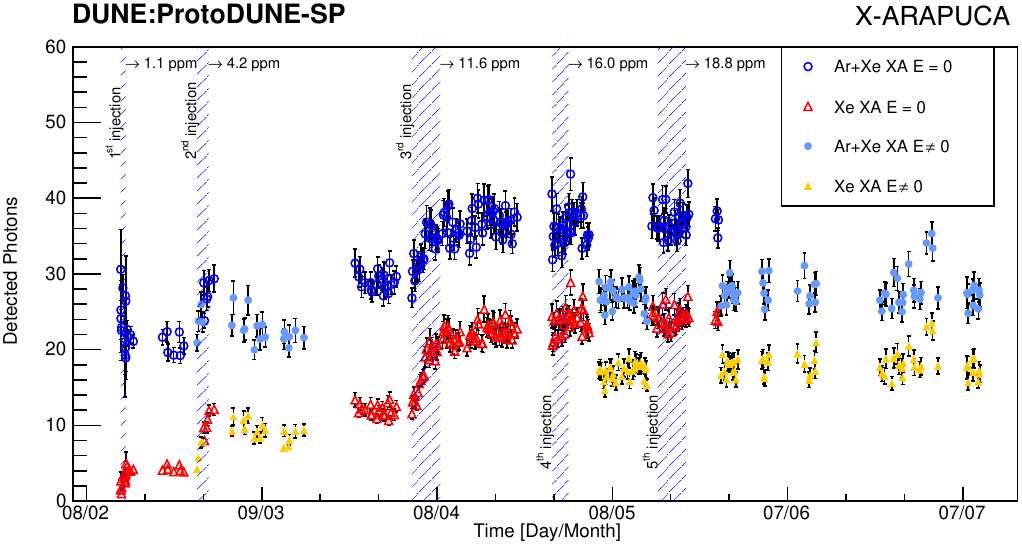}
    \caption{ Light collected by the two X-ARAPUCA modules, in units of detected photons per trigger. Uncertainty bars are derived from statistical uncertainties evaluated on the spread of their distributions. 
    Shaded areas represent xenon injection periods. The amount of collected light increases at each doping. Abrupts steps in the number of photons away from the injection periods correspond to the activation of the HV system of ProtoDUNE-SP and hence the presence of a non-zero electric field. Correspondingly, empty markers are results with no electric field, whereas filled markers correspond to runs with electric field (500 V/cm). }
    \label{fig:light}
\end{figure}

After the single photoelectron calibration, the absolute number of photons detected by the two X-ARAPUCA detectors can be determined. 
Figure~\ref{fig:light} shows that for both X-ARAPUCAs this number increases during each injection and remains stable during the monitoring period following the last injection. This trend is evidence to the effectiveness of energy transfer described in figure~\ref{fig:xe-transition}, especially in the presence of N$_2$. Indeed, the light that was lost after the pollution event appears to be recovered once xenon starts competing with N$_2$-induced quenching. We note that, while it is widely reported in literature that xenon effects on light emission extend up to few hundreds~ppm concentration (e.g.~\cite{Akimov_2019,Neumeier_2015_150nm,Buzulutskov_2017}), in this particular case (i.e., with this Ar-N$_2$-Xe mixture and with these detectors) the increase appears to flatten out at the level of around 16~ppm of xenon. While it is possible that this equilibrium situation is due to surviving nitrogen quenching, more data at higher xenon concentrations would have been required to draw a definitive conclusion on this aspect. Data collected in the two months following the last xenon injection continue this trend, indicating stability of the xenon doping effect on this timescale. 

The presence of an electric field is well known to reduce the number of photons produced as a result of the ionization-recombination process; this effect is evident in figure~\ref{fig:light}.

$\,\,$

\begin{figure}[ht]
\centering
    \includegraphics[width=0.99\textwidth,keepaspectratio]{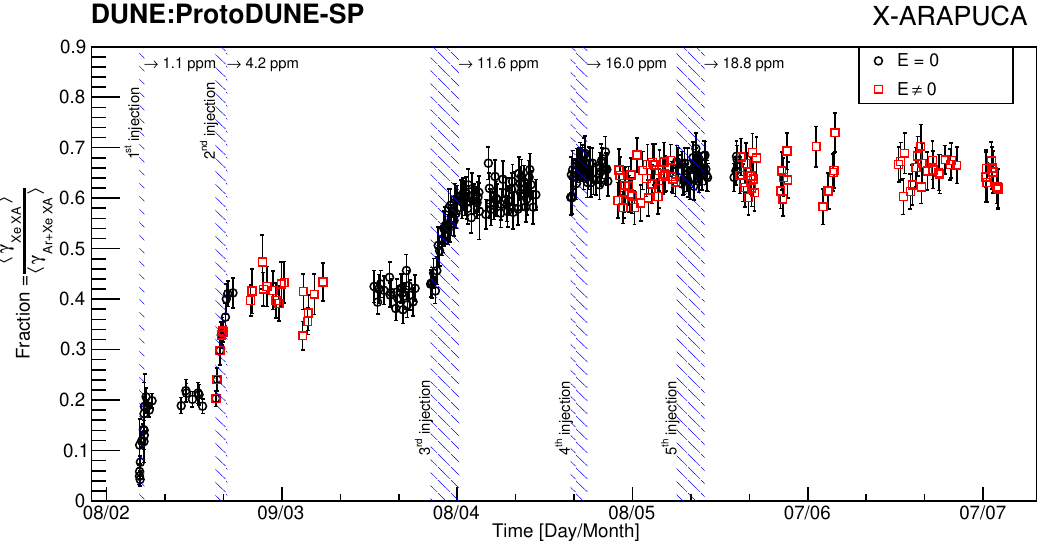}    
    \caption{Fraction of light collected by the xenon-only sensitive X-ARAPUCA: $\frac{\text{Xe}}{\text{Ar+Xe}}$. The ratio increases with the doping and reaches a plateau around 0.65 for xenon concentration greater than \SI{16.0}{ppm}. The red points correspond to data collected with the nominal TPC electric field, while black points refer to data with no electric field. Shaded areas indicate xenon injections.}
    \label{fig:ratio}
\end{figure}

A quantitative estimate of the amount of argon excitation shifted to xenon can be obtained from the observation of the light collected by the two X-ARAPUCA detectors. In this particular case, one calculates the ratio between the xenon light and the total light detected for each run, that is, the ratio of the average light collected by the Xe X-ARAPUCA (only sensitive to xenon, see section~\ref{sec:xArapuca_telescope}) to the average light seen by the Ar+Xe X-ARAPUCA (sensitive to the total light):
\begin{equation}
\text{Fraction}  \equiv \frac{\langle \gamma_{\text{Xe XA}} \rangle}{\langle \gamma_{\text{Ar+Xe XA}} \rangle} = \epsilon \frac{\text{Xe light}}{\text{Ar light} + \text{Xe light}}
\label{eq:ratio}
\end{equation}
Figure~\ref{fig:ratio} shows that this ratio increases as a function of time. 
In particular, as with the total number of photons, this ratio increases with each doping and reaches a more stable value at around~0.65 at \SI{16.0}{ppm}; accounting for the $\epsilon \sim 80\%$ transparency of the fused silica window of the Xe X-ARAPUCA, the ratio becomes~0.81. It should be noted that this is not the final fraction of converted light, since a precise knowledge of the different conversion efficiency of pTP at 127~nm and 178~nm is not available yet. However, the stability of the value since the second to last doping suggests the potential explanation that at this point the \textit{slow} component of light is dominated by photons coming from Xe$_2^*$ dimers. It can also be noted that the Fraction quantity is not expected to be sensitive to electric field. In fact, the presence of the electric field would impact both numerator and denominator of equation~\ref{eq:ratio} with the same scale factor. The reduction of e$^-$ - Ar$^+$ recombination, due to the electric field, would result into a lower amount of Ar$_2^*$ dimers created and thus to a proportional reduction of energy transfer to xenon dimers.

Indeed, the trends of the datasets in figure~\ref{fig:ratio}  with and without the TPC electric field are superimposed. This suggests there is no detectable interference between the electric field presence and the argon-xenon energy transfer process, at least at the level of this measurement in the relatively small region of the detector near the X-ARAPUCAs. 

$\,\,$

\begin{figure}[ht]
\centering
    \includegraphics[width=0.99\textwidth,keepaspectratio]{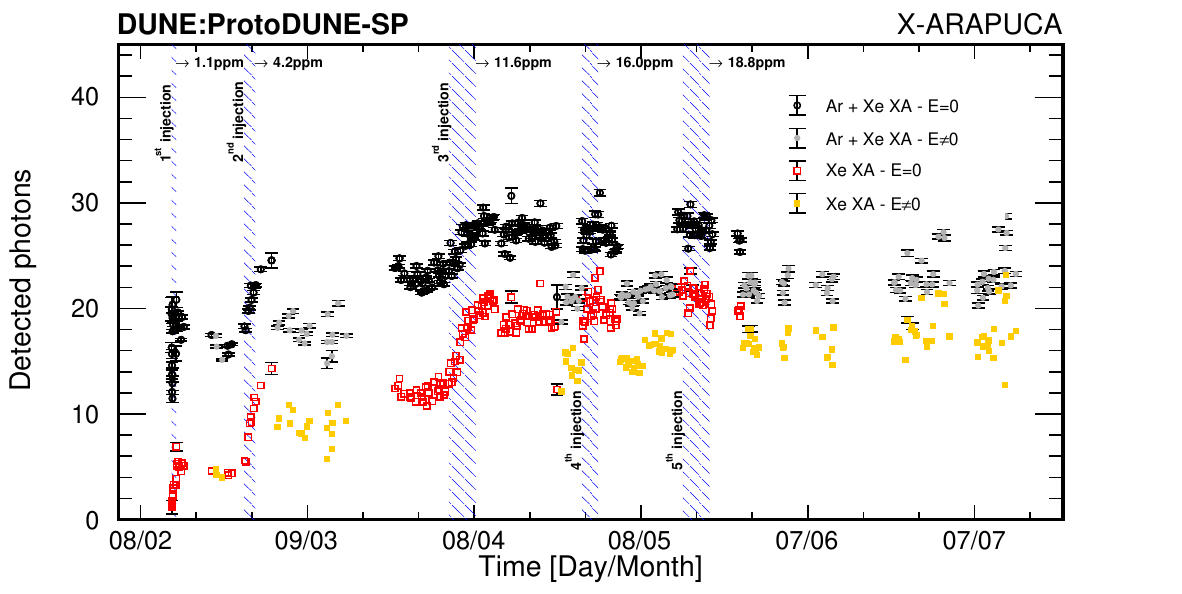}    
    \caption{Time dependence of the mean number of photons in the \textit{slow} light component (detected photons with $t>74$~ns after trigger) detected by the Ar+Xe-light sensitive X-ARAPUCA and by the Xe-light-only sensitive X-ARAPUCA, for runs with and without an external electric field.
    Shaded areas indicate xenon injections.
    Only events with at least three detected photons in the Ar+Xe X-ARAPUCA module are selected. }
    \label{fig:LY_slow}
\end{figure}

Further information about the effect of xenon presence can be extracted by surveying the evolution of the amount of previously defined \textit{fast} and \textit{slow} light components independently, as a function of time.
Figure~\ref{fig:LY_slow} shows the evolution in time of the \textit{slow} light, which now represents the superposition of the residual triplet argon scintillation light and part of the xenon-converted light. 
The start-time value for the separation between the \textit{fast} and \textit{slow} components ($\sim 74$~ns) is set to account for the rise-time of the pulse plus around 3~times the decay-time constant of the argon singlet light. As reported in the literature (e.g.~\cite{Buzulutskov_2017, Acciarri_2010_N2}), this constant should be 6-7~ns, however the convolution with the time response of the ProtoDUNE-SP light detectors~\cite{Abi_2020_PDSPperf} results in a fitted singlet decay-time constant of 13-14~ns.

The number of photons from the \textit{slow} component is shown to increase with xenon concentration, with a trend quite similar to that of the overall light output produced in figure~\ref{fig:light}. This is expected and consistent with the fact that the energy transfer process involves the argon long-lived triplet state (see section~\ref{sec:xe_doping}).
The trends observed for the two X-ARAPUCA are qualitatively quite similar and can be attributed entirely to 178~nm xenon scintillation light, which is further evidence of the physical origin of the light increase.

Figure~\ref{fig:LY_fast} shows the evolution of the \textit{fast} light component with time. The plot shows a very quick drop of this light during the first doping period, followed by a small,  stable output throughout the rest of the run. 

The rapid drop of the \textit{fast} light observed at the beginning of the doping was unexpected. It cannot be explained a priori by the xenon energy transfer process, as the argon singlet decay time ($\tau_s = 6$~ns) is much shorter that the time required for the Ar$_2^*$ - Xe interaction to take place.
However, there are studies in the literature~\cite{Neumeier_2015} that report direct absorption of the argon light by xenon, ascribed to the absorption spectrum of xenon partially overlapping with the 127~nm scintillation peak of argon, which has a FWHM of around 10~nm. There, the absorption process seems to be saturating at the lowest concentrations of xenon, which is consistent with our observations.
If that is the case, the residual \textit{fast} component detected in the fully sensitive X-ARAPUCA can be ascribed to the singlet argon photons surviving absorption.

That said, going back to figure~\ref{fig:wf_comp}, we notice that the extinction effect is very clearly visible even in the xenon-only sensitive X-ARAPUCA (bottom panel), despite the fact that it is expected to be due only to argon singlet de-excitation.

Indeed, figures~\ref{fig:wf_comp} and~\ref{fig:LY_fast} show that despite the process described above and the fact that the xenon-only sensitive X-ARAPUCA is completely opaque to 127~nm photons, on average one/two photons in the \textit{fast} component (i.e. within $\sim$74~ns from trigger) are still detected by this detector, for any xenon concentration. 
Their origin is not obvious, however there are possible sources not related to the primary scintillation process: Cherenkov emission from cosmic rays secondary particles crossing the device entrance window; wavelength-shifted light escaping other PDS modules and entering the device inner volume; or spurious events inside the inner volume.  Further data and a more refined model of the energy transfer mechanism (e.g. evaluation of $\tau_{AX}$ and $\tau_{XX}$) are necessary, in order to understand if, for example, the xenon light obtained in the transfer process can partially contribute to this component. 

Strictly interpreted, this analysis of the X-ARAPUCA data is valid as a measurement of the xenon effect on argon scintillation light for a limited dataset of cosmic muons over a limited angular range and in a limited region of the ProtoDUNE-SP detector. Nonetheless, it provides a general confirmation of the argon-xenon energy transfer hypothesis as an explanation of the observed phenomenon, and it demonstrates stability of the effect for a period of at least two months following the final injection.

These data do leave some open questions, however, especially concerning the \textit{fast} light component detected by the X-ARAPUCAs. In order to better understand this aspect, more data is required. Soon after the ProtoDUNE-SP xenon run, a similar campaign was carried out with the other DUNE prototype at CERN, ProtoDUNE Dual Phase (DP). The photon detection system of this second detector was composed of thirty six 8-inch cryogenic model R5912-02MOD photomultiplier tubes from Hamamatsu~\cite{DuneDP_IDR}. Measurements were initially made with pure LAr, whose level was then allowed to drop by evaporation, and then partially re-filled with 230~t of LAr+N$_2$+Xe transferred from ProtoDUNE-SP. This produced a mixture with 5.8~ppmv (ppm in volume) Xe and 2.4~ppmv N$_2$. Measurements were also taken after further injections of N$_2$ to bring the levels to 3.4~ppmv and 5.3~ppmv. ProtoDUNE-DP has a different technology and a larger drift distance of 6~m, thus allowing both detectors to probe a wide range of drift paths. It is interesting in particular to note that ProtoDUNE-DP observed a similar drop in the \textit{fast} component of argon light, after the injection of xenon. Results from the Dual Phase data-analysis have been published~\cite{SotoOton:2812306,pddp_pds} and a joint analysis of datasets from Single and Dual Phase detectors is anticipated.

\section{Analysis of the ProtoDUNE-SP PDS during the xenon-doping periods}
\label{sec:analysis_protodune_pds}

Although it is not optimized to differentiate between the light produced by the liquid argon and xenon, the standard ProtoDUNE-SP photon detection system provides an independent measurement, including data from before the nitrogen contamination event~\cite{Abi_2020_PDSPperf}. 

The dataset discussed in this section is divided into multiple epochs: a period before xenon doping and nitrogen contamination, labelled as the first ProtoDUNE-SP run; a period after the first run, with only nitrogen contamination present in the drift volume; and a xenon doping period, where xenon was injected over a period of few months.
Throughout these data taking periods, the TPC electric field settings varied from zero to the nominal setting (500~V/cm), significantly changing the total amount of scintillation light available for detection.
All the following ProtoDUNE PDS studies use light collected from through-going cosmic-ray muons selected in coincidence with the cosmic ray tagger.

\subsection{Triggering, data selection, and collected light}
Triggering in ProtoDUNE-SP relies on the central DAQ and requires coordination between two or more subsystems.
For the ProtoDUNE-SP PDS, two major triggering schemes exist, both depending on a coincidence between the upstream and downstream modules of the CRT. 
The trigger coincidence window length, pre-scaling, and trigger mask varied throughout the run.

If TPC track information is available, an upstream and downstream CRT coincidence is correlated with through-going tracks, allowing a comparison of the orientation of the track, as reconstructed by the TPC, to the vector that intersects the strip hits in both triggered CRT modules.
Single tracks are selected for later analysis if they meet the TPC reconstruction and selection criteria, have a viable trigger and light signals, and if they pass a quality cut of $(\cos{\theta} > 0.999)$, i.e., if the angle $\theta$ between the track from TPC and this vector has a deviation of less than a degree.
If the TPC information is not available, a selection is made based on matching distinct PDS coincidences across APAs, with coincident strip hits in the upstream and downstream CRT modules, requiring at least two photon detectors in two different APAs within a time coincidence of 13~$\mu$s.

The light collected from the selected sample is summed across a single detector and assigned a \textit{radial distance}, which is defined as the straight line distance from the photon detector to the track, when they are in the same XY-plane (vertical plane normal to APAs).
A Gaussian or Poissonian fit  to the distribution of the collected light at each~centimeter of radial distance is performed, to obtain the most probable value. This represents the expected amount of light observed from a passing muon at a given radial distance; the choice of function used for the fit is determined by the bin statistics. Details concerning the PDS calibration and performance can be found in ref.~\cite{Abi_2020_PDSPperf}.

\begin{figure}[h]
    \centering
    \includegraphics[width=0.99\textwidth]{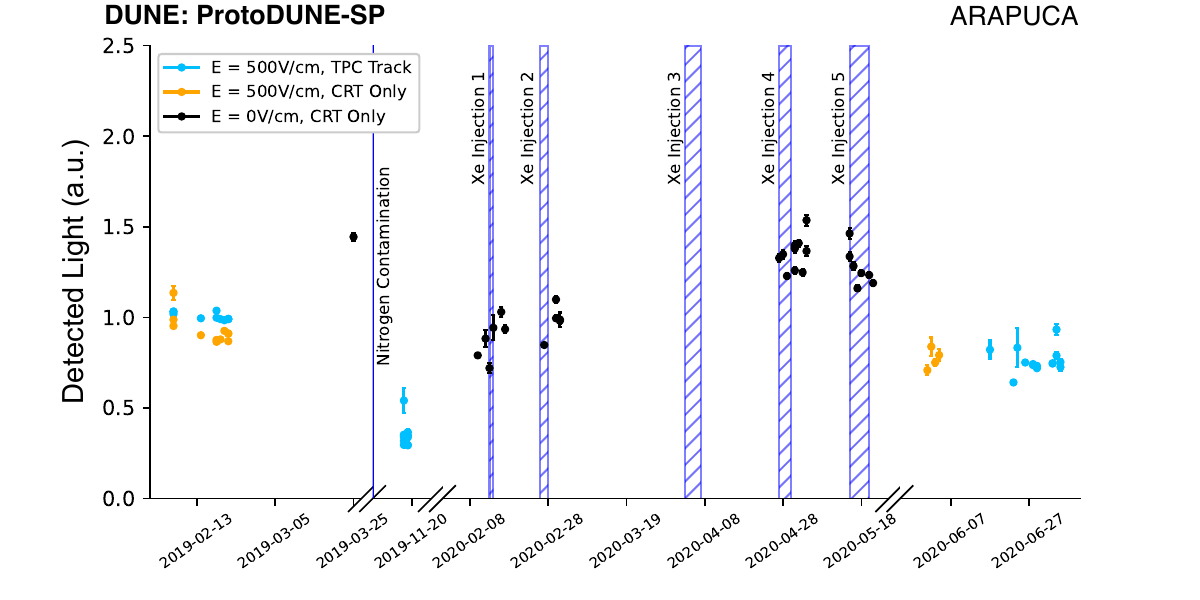}
\caption{Average light signal detected in the non-beam side standard ARAPUCA, across the nitrogen contamination and xenon doping period. Blue lines indicate changes in the scintillation medium through nitrogen contamination or xenon injection. Variations in detected light with run conditions are clear, both for TPC-reconstructed tracks with full electric field (blue) and for CRT-triggered tracks without a reconstruction match with field on (orange) and off (black).\label{fig:triglightblara}}
\end{figure}

An analysis of the average collected light as a function of time and with different trigger periods is shown for a PDS Standard ARAPUCA, on the non-beam side, in figure~\ref{fig:triglightblara}. The doping periods are indicated by the vertical shaded blue areas in the plot. Tracks were selected between 100 and 200~cm from the ARAPUCA module to minimize the effects of run-to-run variations in trigger geometry. Data from a selection of high-statistics runs with the TPC electric field at nominal (500 V/cm) and off are compared. Some runs with nominal field were not matched to TPC reconstruction, instead relying on the CRT for tracking information, to emulate the field-off runs. The mean integrated waveform across selected events within a run is normalized to the average signal from TPC-matched tracks during the pure-LAr phase in February 2019.

Data from the ARAPUCA show a qualitatively consistent behavior when compared with the X-ARAPUCA results described in the previous section. 
The average amount of light detected in the ProtoDUNE PDS drops after the nitrogen contamination and increases, in steps, with each additional doping with xenon. 
Data collected with and without the TPC electric field consistently show two compatible trends of increase, due to the different available total amount of scintillation light. 
TPC-on and TPC-off data are indicated by different colored points in figure~\ref{fig:triglightblara}.

It should be noted that the average amount of light at the end of the xenon doping campaign is roughly comparable to that from before the nitrogen contamination, indicating that, as with the X-ARAPUCA data, the injection of xenon compensates the negative effect of contamination.
As it will be shown in the next section, this ``recovery'' of light is not uniform with distance, but, beneficially, it becomes relatively more significant when the light source is farther from the detector.

The comparison between X-ARAPUCA and ARAPUCA data is more straightforward, given the similar technology. It should also be noted that ARAPUCA were the detectors showing the highest efficiency in ProtoDUNE SP, among those installed (see ref.~\cite{Abi_2020_PDSPperf}), leading to the improvement of the technology and its selection for installation in DUNE . However, for this campaign, data from the other photon detectors in ProtoDUNE-SP, DSLG and DCLG, were analyzed as well. We note that the behaviour observed in figure~\ref{fig:triglightblara} is consistent across these other light detectors, despite differences in the absolute number of collected photons in these different technologies, further supporting the qualitative conclusions of this study.

\subsection{Light recovery due to xenon injection}

As described in the previous section, the amount of light collected by the ProtoDUNE PDS and the changes in the typical light-pulse profiles (waveforms) can supply critical information about how the injected xenon significantly alters the character of the scintillation light produced in the detector. 
The large sample of through-going cosmic muon tracks from ProtoDUNE-SP allows for the construction of \textit{attenuation curves,} that track the number of detected photons as a function of the radial distance. 
Most of the plots in this section show data collected by the non-beam side ARAPUCA, i.e. the one in the Left TPC, with respect to the beam direction (see section~\ref{sec:intro}). 
The phrase \textit{Pure LAr} in the plots legends indicates data from the period before the nitrogen contamination.

\begin{figure}[htbp]
    \centering 
\subfloat[]{
\includegraphics[width=0.45\textwidth]{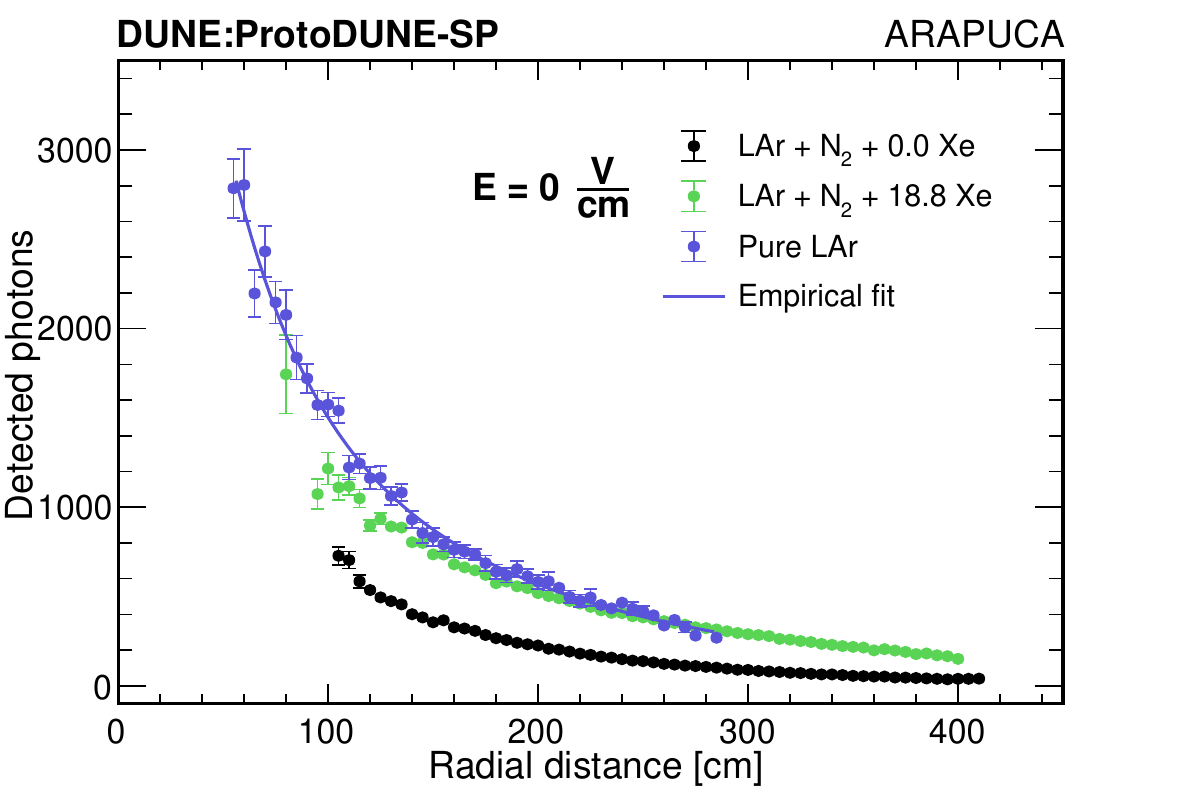}
}
\subfloat[]{
\includegraphics[width=0.45\textwidth]{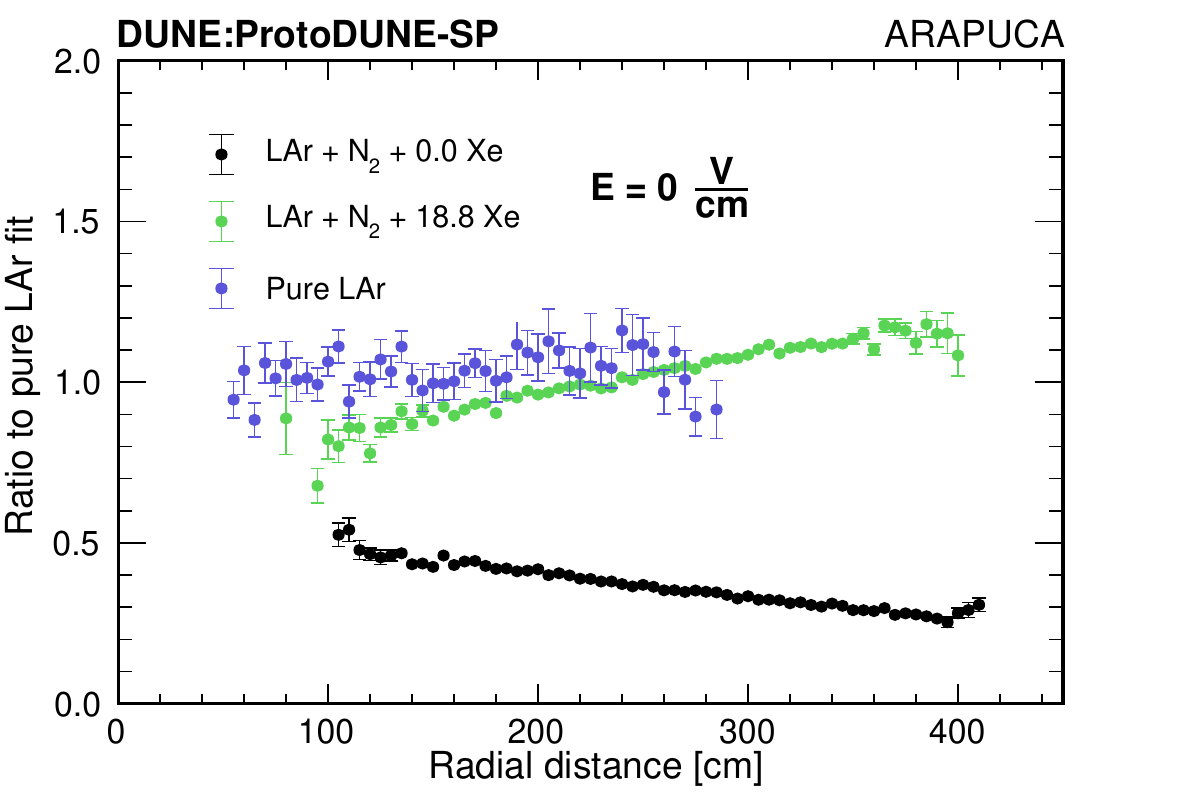}
}

\subfloat[]{
\includegraphics[width=0.45\textwidth]{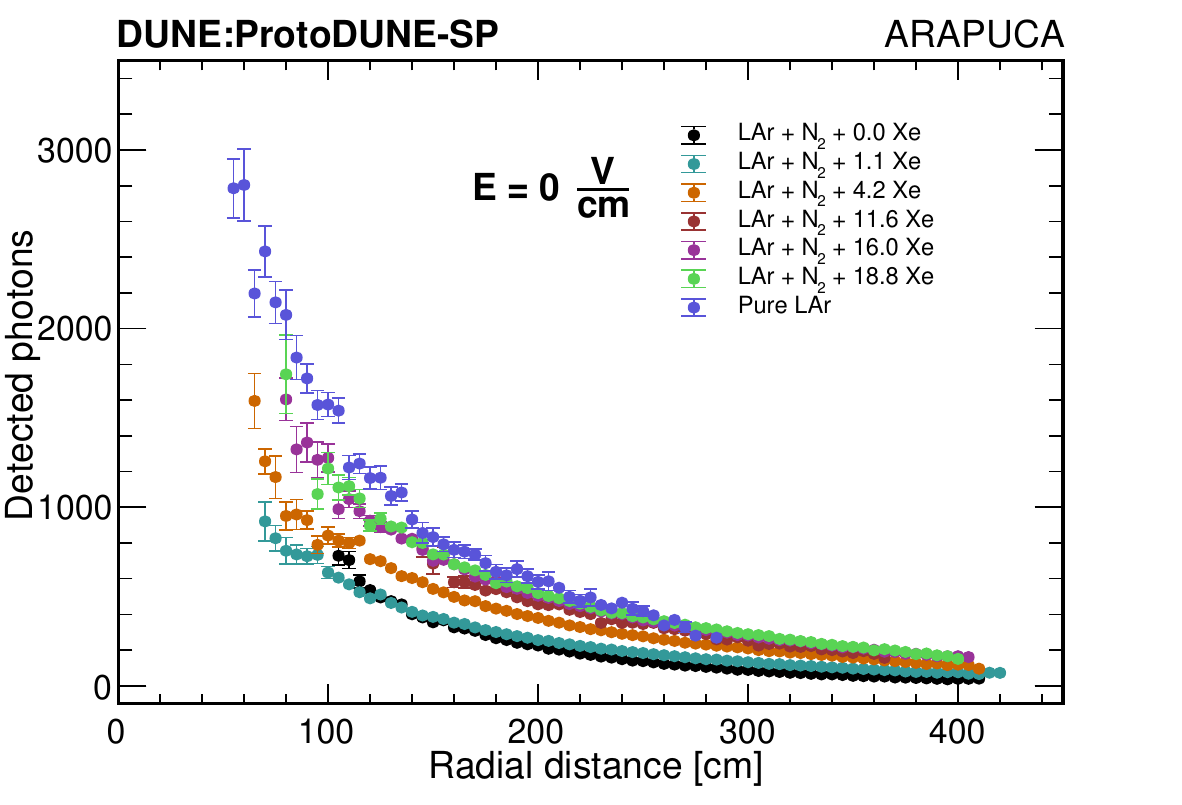}
}
\subfloat[]{
\includegraphics[width=0.45\textwidth]{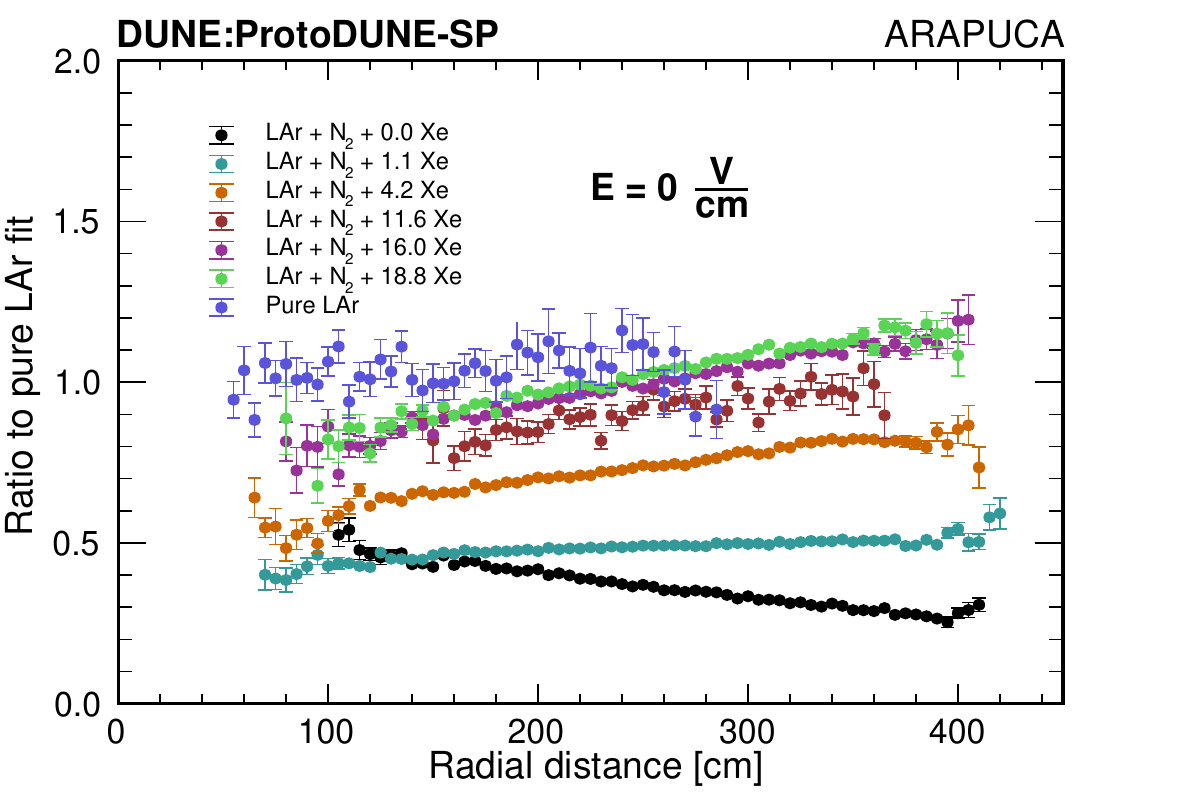}
}

\subfloat[]{ 
\includegraphics[width=0.45\textwidth]{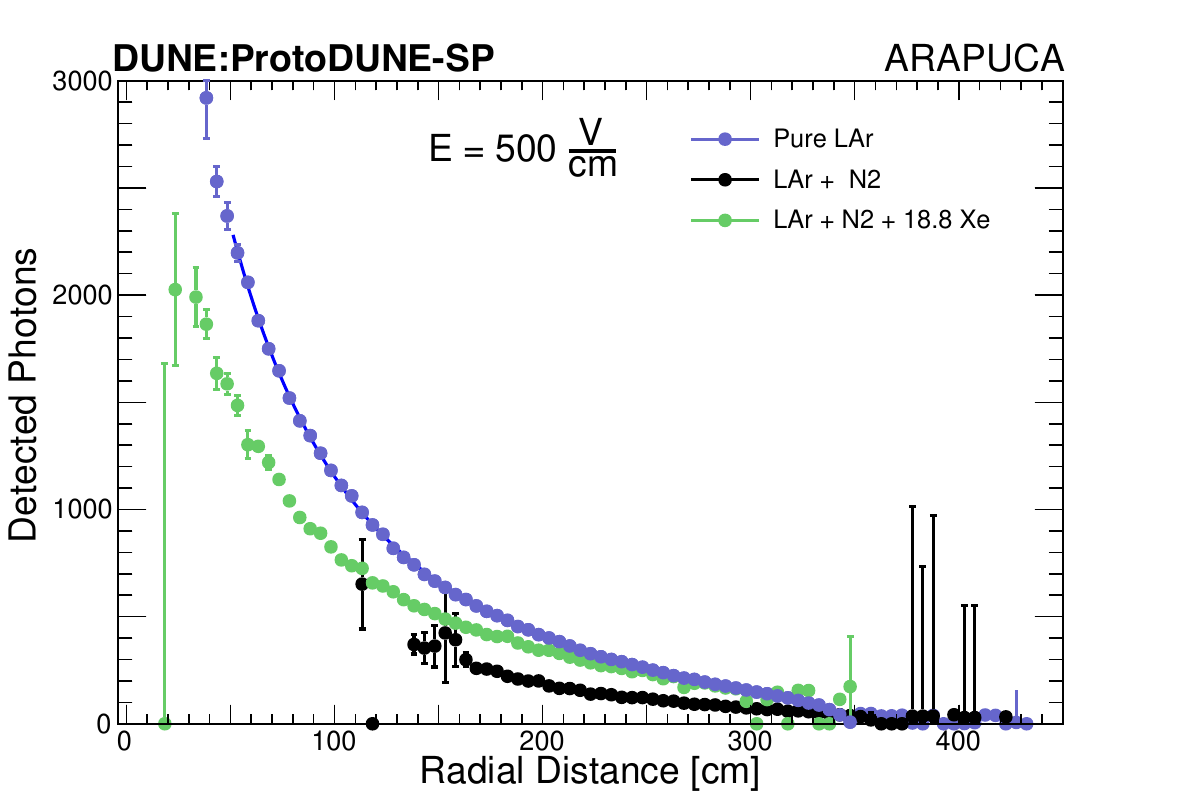}
}
\subfloat[]{
\includegraphics[width=0.45\textwidth]{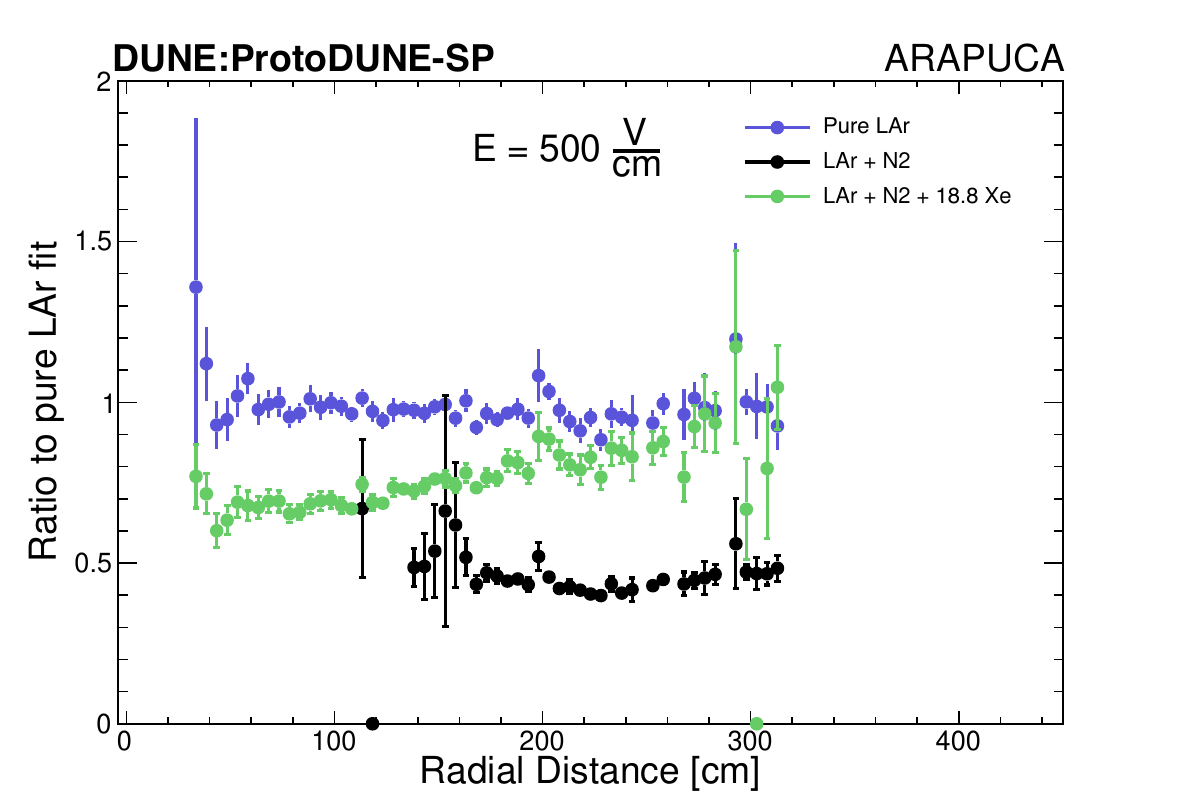}
}
\caption{\label{fig:06_LightRecoveryAra} Light recovery demonstrated through attenuation curves after xenon injection with the non-beam side PDS ARAPUCA. The left column of plots shows the collected light versus radial distance, while the right column shows the ratio of collected light with nitrogen and with nitrogen+xenon, relative to the fitted pure-LAr conditions. The empirical functional form $a\exp(-r/l_1) + b\exp(-r/l_2)$ is fitted to pure-LAr data, with $a=2200 \pm 400 ~(1144 \pm 52)$ Detected Photons, $b=6000 \pm 1000 ~(2662 \pm 115)$ Detected Photons, $l_1= 140 \pm 10~(161 \pm 5)$~cm, $l_2= 37 \pm 7~(39 \pm 2)$~cm without (with) electric field.
The top and bottom rows of plots show the measurement made without and with TPC electric field, respectively. The middle plots detail the gradual increase of collected light with increasing xenon concentration, with no drift field.}
\end{figure}

Figure~\ref{fig:06_LightRecoveryAra} shows the amount of light collected by the non-beam side ARAPUCA as a function of the radial distance of the light from the sensor, for different periods. Panels on the left column show the  number of detected photons, whereas panels on the right provide the ratio of light with respect to the fitted period with \textit{Pure LAr}. For clarity, the top and bottom rows include data only for the non-doped and highest xenon-doping level datasets, the middle row includes all five dopant levels. 
Left-column panels clearly show the decreasing amount of collected light, as a function of the event distance from the sensor, not corrected for solid-angle. 
The effects of contaminant and dopant are more easily appreciated in the ratios provided in the right-hand column.
Panels (b),(d),(f) clearly show how nitrogen effectively lowers argon light production uniformly across the TPC drift distance. It is also apparent that all panels demonstrate the increase in collected light due to xenon doping.
Panels (c) and (d) show all five xenon doping levels results, with the separation between the injections reducing as the concentration of xenon goes up. 

Comparison between panels (b) and (f) shows similar qualitative behavior with and without the presence of the TPC electric field, again indicating no detectable interference between the xenon doping and the TPC operation.
The right-column panels also confirm that the amount of collected light recovered after the doping is higher far away from the photodetectors and lower close to the sensors, as compared to the pure argon case. Such dependence on distance is due to the larger Rayleigh scattering length of 178~nm photons in LAr, with respect to that of 127~nm light. The effect is  evident even for the lowest xenon injection level investigated, whereas no such change in slope is detected in the Ar-N$_2$ case, as expected.
This effect can partially mitigate the intrinsic non-uniformity of the ProtoDUNE PDS coverage, which is installed only in the proximity of the TPC anode (i.e. embedded inside the APAs).

\begin{figure}[htbp]
\centering
\subfloat[]{\includegraphics[width=0.49\textwidth]{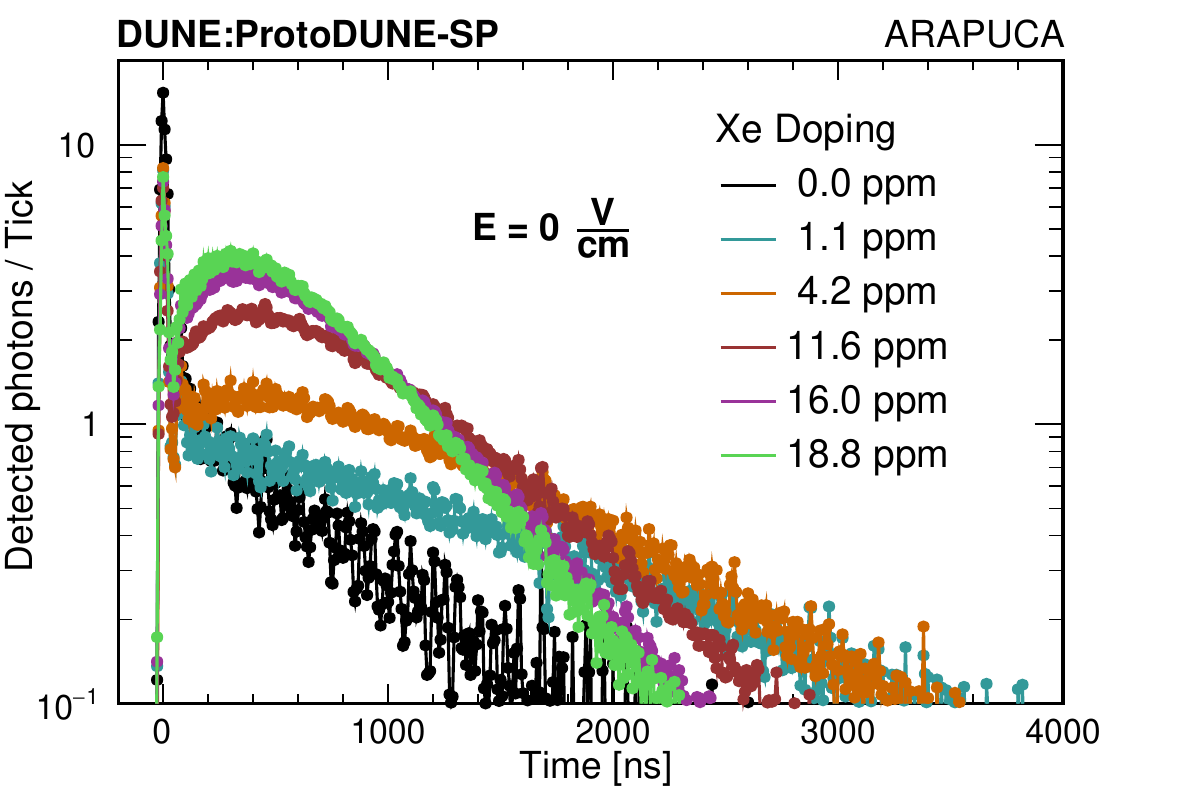}}
\subfloat[]{\includegraphics[width=0.49\textwidth]{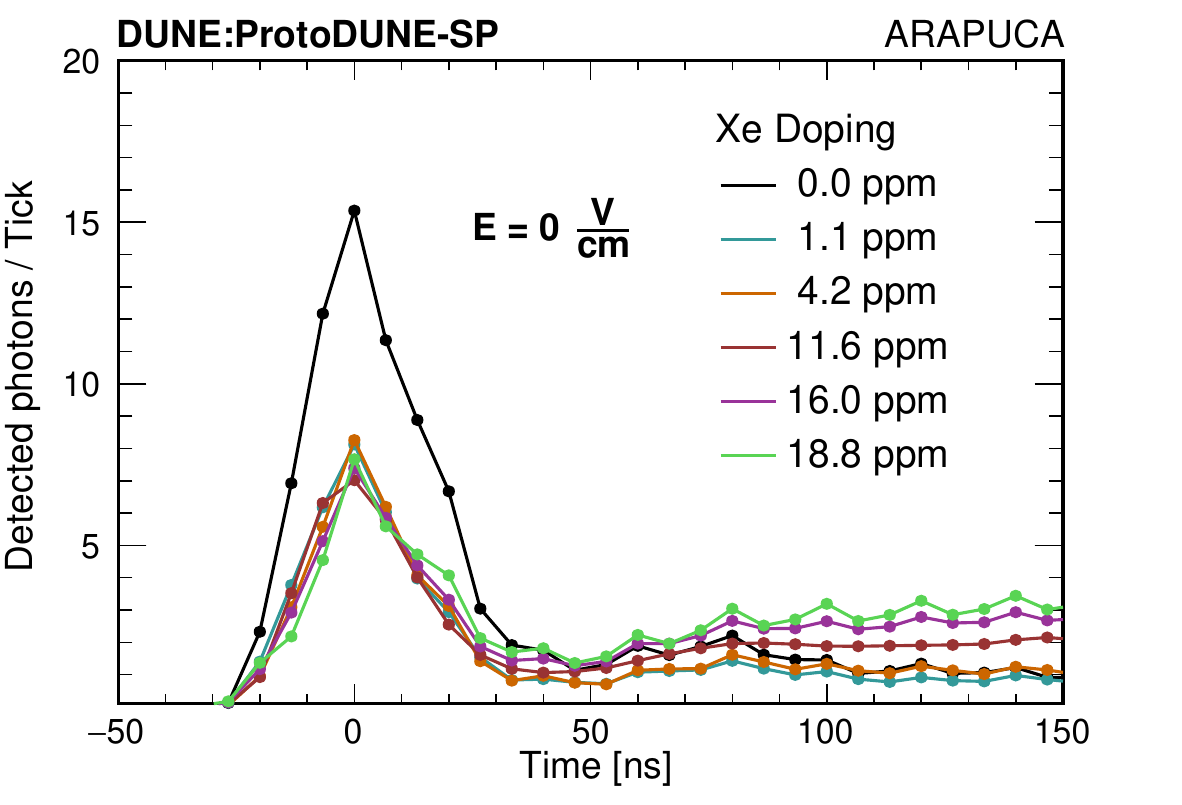}}

\subfloat[]{\includegraphics[width=0.49\textwidth]{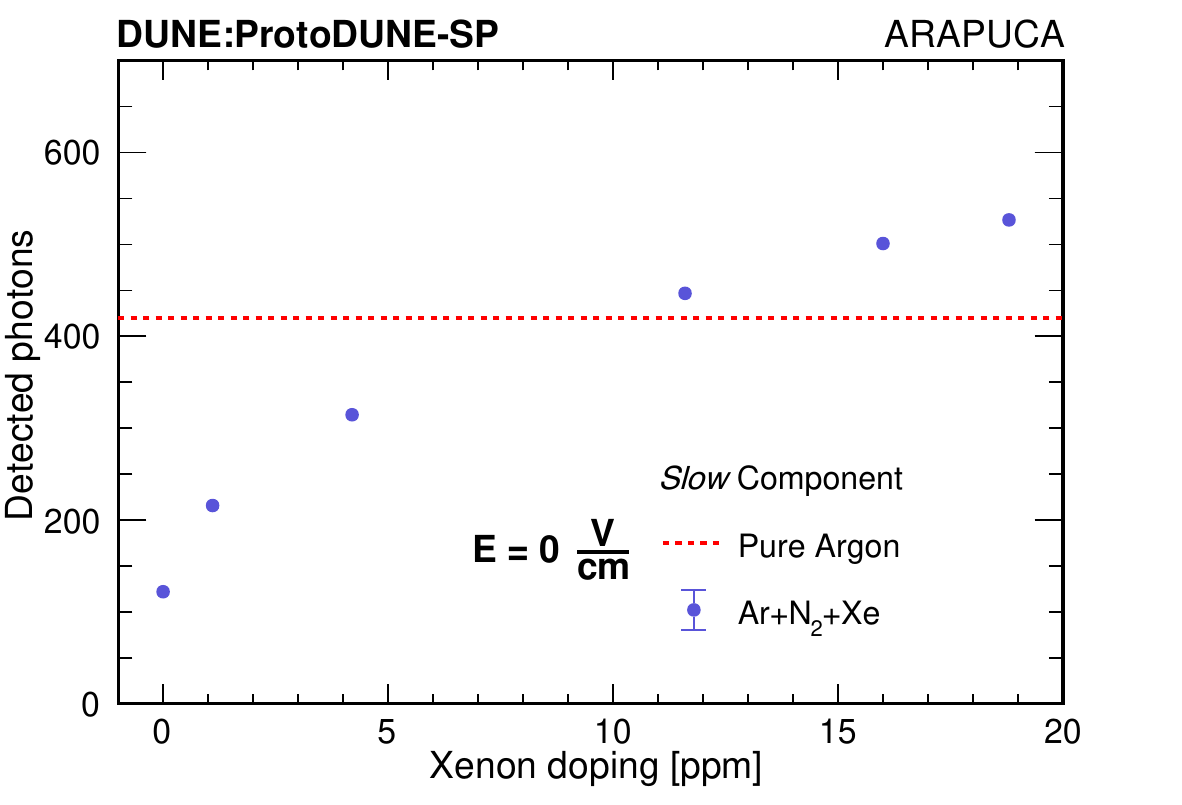}}
\subfloat[]{\includegraphics[width=0.49\textwidth]{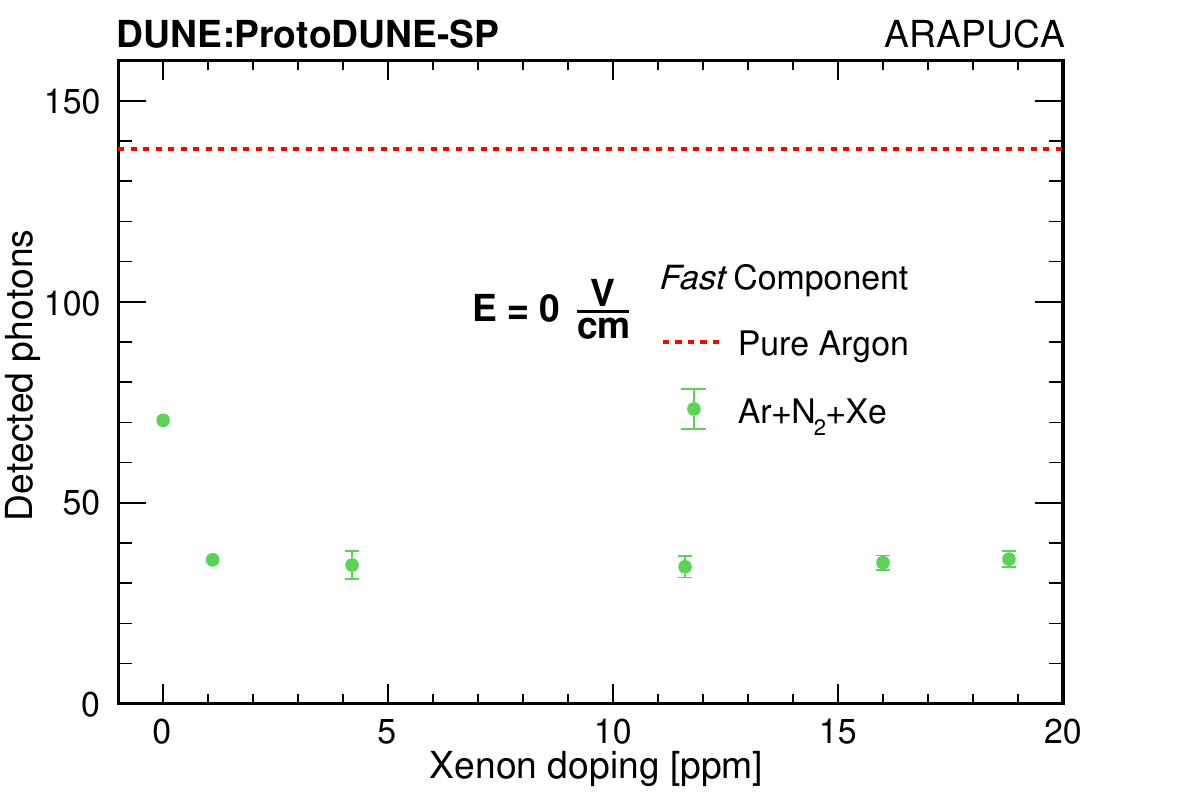}}

\subfloat[]{\includegraphics[width=0.49\textwidth]{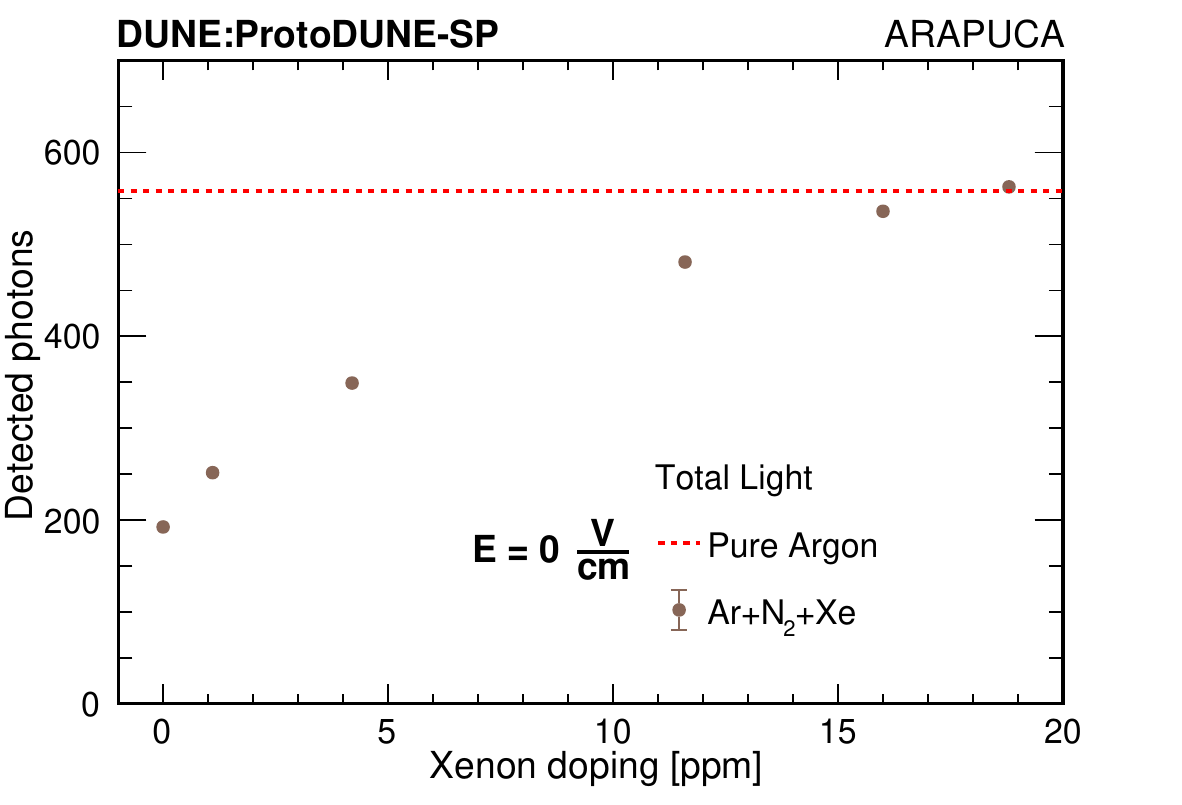}}
\caption{\label{fig:06_XeDopWave} Data from ARAPUCA on the non-beam side TPC. Top row (a, b): deconvolved waveforms, changing in shape with increasing concentration of xenon; ``0~ppm'' data refer to argon polluted with nitrogen. Middle row: evolution of the \textit{slow} (c) and \textit{fast} (d) light components as a function of xenon concentration, in the nitrogen contaminated scintillation medium. Bottom row: evolution of the total detected light (e) as a function of xenon concentration, in the nitrogen contaminated scintillation medium.}
\end{figure}

Figure~\ref{fig:06_XeDopWave} provides more detail of the change in time profile and light output increase measured by the ARAPUCAs. 
Events used in these plots are a subgroup of all the events used for the distributions in panels (c) and (d) for Figure~\ref{fig:06_LightRecoveryAra}. 
The selection was made using cosmic ray tracks with a mean radial distance of about 250~cm, with a standard deviation of about 30~cm. 
The selection in a relatively narrow range is needed because the waveform shape and integrals change with the distance of the event from the detector, due to the differences between the argon and xenon light propagation seen in figure~\ref{fig:06_LightRecoveryAra}.
 
Panels (a) and (b) show that, as the concentration of xenon increases, the portion of waveform corresponding to the \textit{slow}\footnote{For the analysis of the PDS data, the same definitions of \textit{slow} and \textit{fast} component of the scintillation light still hold, in terms of intervals of integration, as given in section~\ref{sec:analysis_xArapuca}.} component of the characteristic argon pulse is increased by at least a factor of five. On the other hand, the characteristic argon \textit{fast} component is significantly reduced, by around a factor~2, already at 1~ppm of xenon, and then it remains stable throughout the subsequent doping steps. 
These trends are consistent with those obtained from the analysis of the X-ARAPUCA data, using a different detector and dataset.
Panels (c), (d), and (e) in figure~\ref{fig:06_XeDopWave} summarise the changes in the average number of detected photons across the full xenon doping period for the \textit{slow} and \textit{fast} components of the scintillation light, as well as for the total collected light.

Xenon injection affects the light recovery both at the level of the scintillation process and light propagation, given the different wavelength. The overall effect averaged on the radial distance shows a final amount of the detected light which is about 95\% with respect to the pure LAr case, with a total concentration of 18.8~ppm of xenon.

\begin{figure}[ht]
    \centering
    \includegraphics[width=0.45\textwidth]{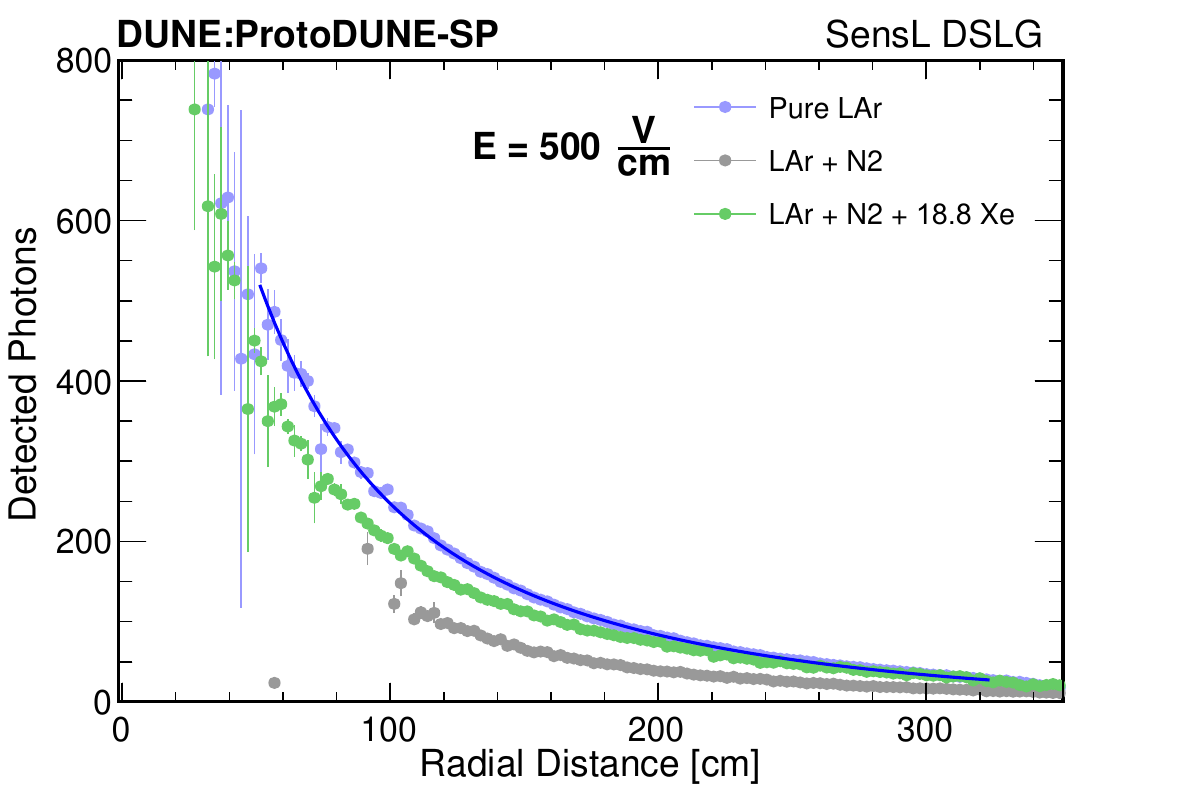}
    \includegraphics[width=0.45\textwidth]{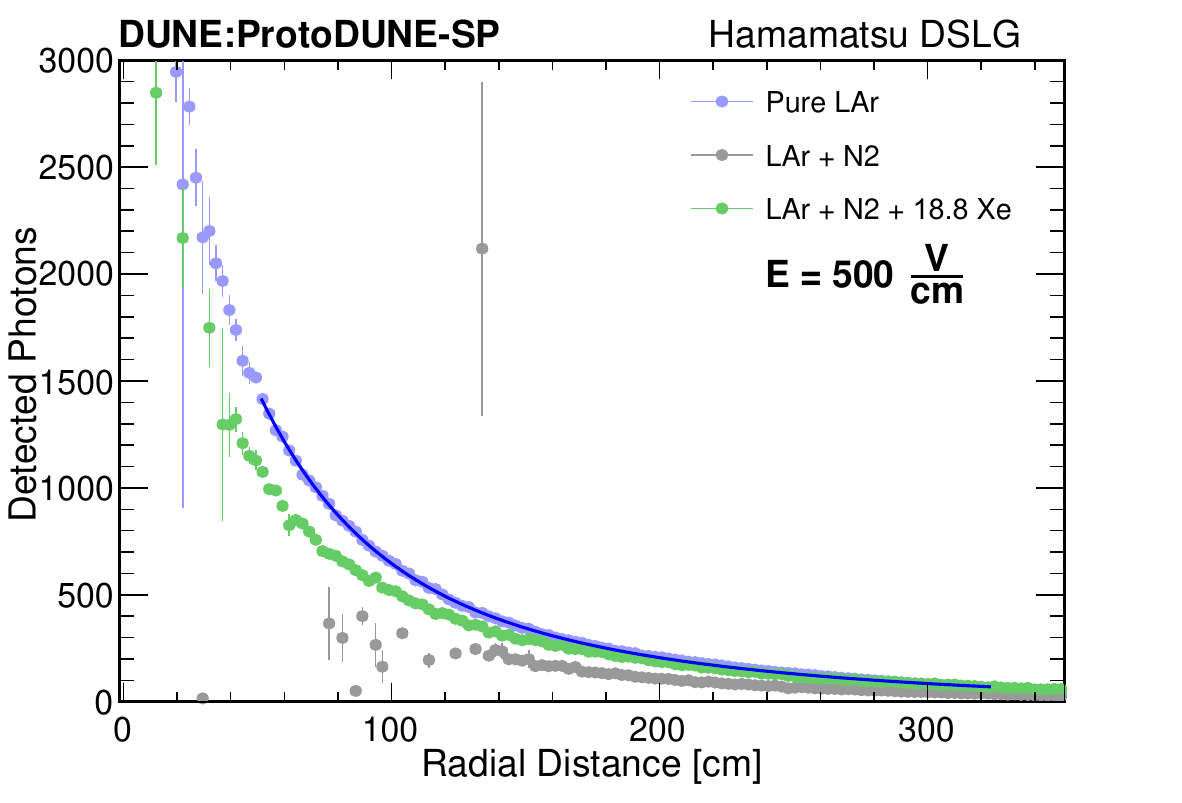}
    
    \includegraphics[width=0.45\textwidth]{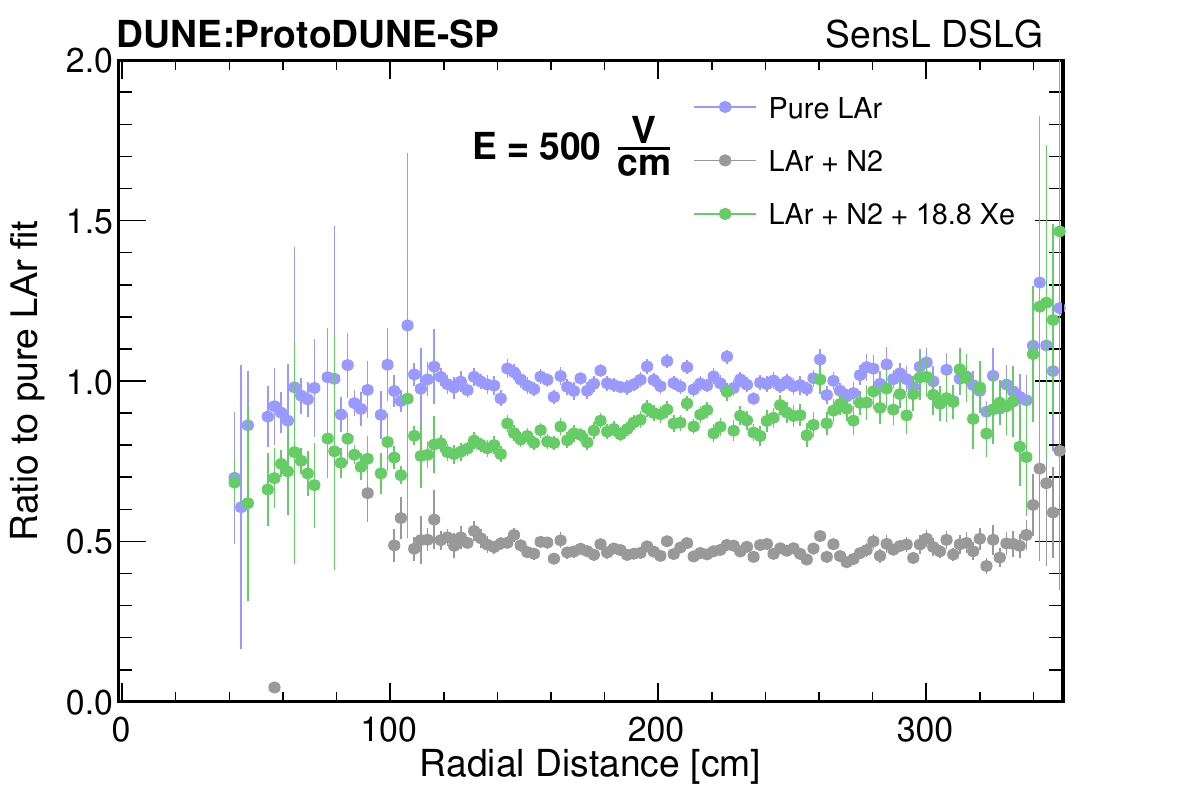}
    \includegraphics[width=0.45\textwidth]{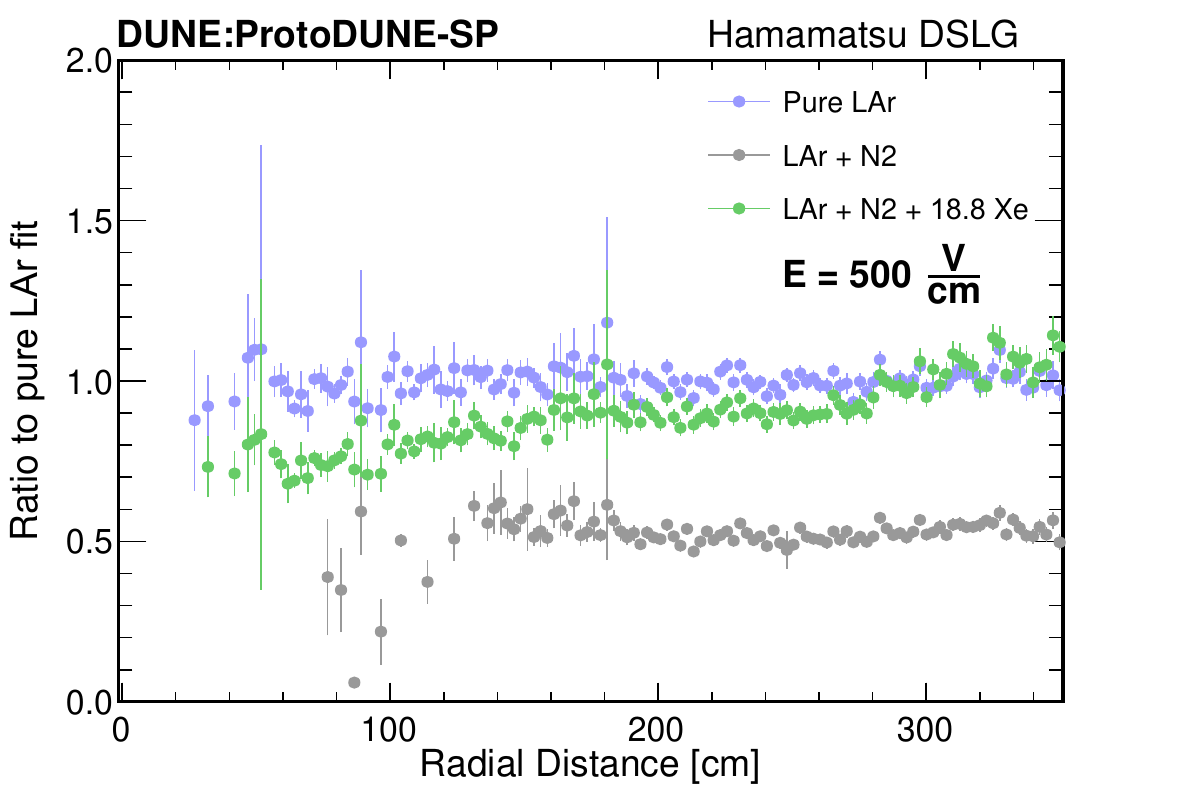}
    \caption{Light recovery as demonstrated through attenuation curves for the non-beam side Double Shift Light Guides, separated by photosensor technology, SensL (left column) and Hamamatsu (right column), with the TPC drift field on. The Ar-N$_2$-Xe data refer to the end of the doping, i.e., to a xenon concentration of 18.8~ppm. The top plots represent the collected light versus radial distance, whereas the bottom plots show the ratio of collected light relative to the fitted pure-LAr conditions. The empirical functional form $a\exp(-r/l_1) + b\exp(-r/l_2)$ is fitted to pure-LAr data, with (\textit{i}) \textit{SensL}: $a=932 \pm 7$ Detected Photons, $b=260 \pm 5$ Detected Photons, $l_1= 38 \pm 1$~cm, $l_2= 163 \pm 2$~cm; (\textit{ii}) \textit{Hamamatsu}: $a=2832 \pm 17$ Detected Photons, $b=888 \pm 4$ Detected Photons, $l_1= 30 \pm 1$~cm, $l_2= 135 \pm 1$~cm.}
    \label{fig:attenuationcurve}
\end{figure}

Figure~\ref{fig:attenuationcurve} presents data from the DSLG detectors; results from modules equipped with SiPMs from SensL (left column) and Hamamatsu (right column) are shown separately. The distributions are qualitatively the same as for the ARAPUCA modules, providing a consistent picture across all the ProtoDUNE-SP photon detectors.
The analysis using the full set of the ProtoDUNE photon detectors confirms on a detector-wide level the results obtained in a more restricted region with the dedicated X-ARAPUCA detectors in section~\ref{sec:analysis_xArapuca}. 
    
\newpage
\section{Charge reconstruction in liquid argon doped with xenon}
\label{sec:charge_analysis}

During the xenon doping run, the operation of the ProtoDUNE-SP TPC was monitored in order to investigate whether the presence of the dopant would affect the charge collection.
A useful monitor of the stability of the ProtoDUNE TPC performance is the so-called \emph{TPC signal strength}.
In ProtoDUNE-SP, the primary contribution to charge deposits in the LAr is ionization from cosmic rays.
The amount of collected charge is evaluated for each collection wire by summing all the calibrated charge deposits, over those regions where the signal is significantly above the noise level for the channel.
The fraction of the originally produced ionization charge actually reaching each collection wire depends on the purity of the LAr, on the voltages applied to the wires and cathode planes, as well as on space charge effects~\cite{Abi_2020_PDSPperf}.
The response of the detector relies on the gain of the electronics modules, which was calibrated with test-charge injections and was stable over the course of the run~\cite{Adams_2020}.

\begin{figure}
\centering
  \includegraphics[width=0.9\textwidth,angle=0,trim=0 0 0 0, clip]{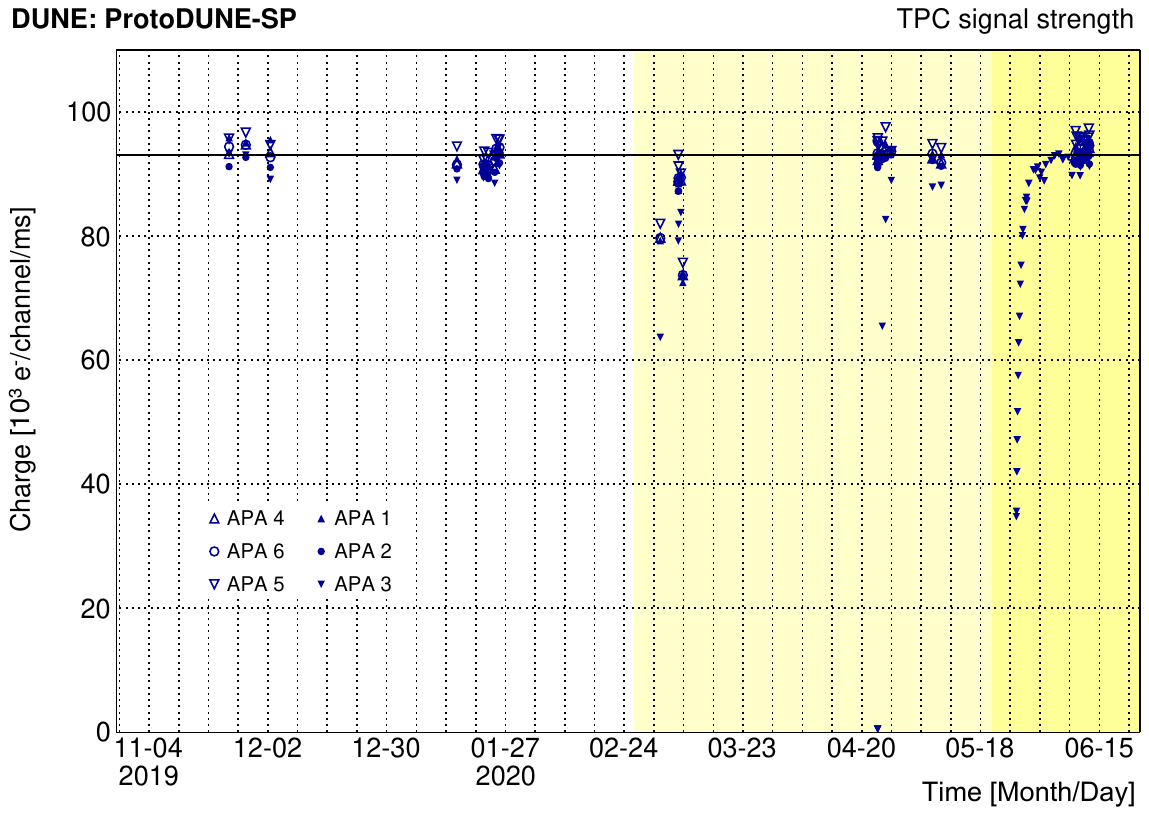}
\caption{Signal strength versus time, before (white), during (light yellow) and after (yellow) the xenon doping campaign. Details concerning the drops in collected charge can be found in text.}
\label{fig:sigstrenxe}
\end{figure}

Figure~\ref{fig:sigstrenxe} shows the TPC signal strength before, during and after the xenon filling, for those periods where APA data were collected with voltages at or near nominal values.
Each point is evaluated by averaging the calibrated charge over all good collection wires in an APA for a few thousand randomly triggered events, with acquisition windows of 3~ms in each event.
The figure includes a line at 93000~e$^-$/channel/ms, which is the reference value for nominal voltages and high purity.
The first drop in signal strength on all APAs during the first doping period is mainly due to the different electric field at which the TPC was operated, ranging from \SI{250}{\V\per\cm} to \SI{500}{\V\per\cm} (nominal voltage). Later low-charge data are only related to APA3, which was suffering biasing issues~\cite{Abi_2020_PDSPperf}.

The xenon doping period is highlighted in light yellow, whereas the darker yellow refers to maximal xenon concentration.
One can see that the average TPC signal strength in standard conditions remains at its nominal value before, during and after the xenon doping.

Additional studies should follow, to confirm that xenon doping can be safely used in the DUNE LArTPCs. However, this initial analysis demonstrates that no major show-stopper is present and that, at the level of $\sim$1~kt mass, xenon has no observable effect on the fraction of charge reaching the TPC collection wires.

\section{Conclusion}

Xenon doping of liquid argon is a known technique to increase scintillation output of the medium and to enhance light collection by shifting the photons to a longer wavelength.  Since such an enhancement would be beneficial to the physics program of the DUNE experiment, it was proposed to be a feature of its second far detector module (FD2), and a large scale test was called for. In this paper we described the first large-scale attempt at xenon doping using the 720~t ProtoDUNE-SP detector at CERN. 

The goal was to perform measurements of the scintillation light output, during and after the xenon doping, using the photodetectors that were part of the original ProtoDUNE-SP configuration, sensitive to light throughout the TPC active volume, along with two dedicated X-ARAPUCA detectors installed outside the active volume. Using a filter, one of the X-ARAPUCAs was configured to be only sensitive to xenon light, whereas the second was sensitive also to argon light.

However, before the test began, ProtoDUNE-SP suffered an accidental leak resulting in nitrogen contamination at the few~ppm level that significantly reduced the amount of light reaching the photosensors. This became a possibility to study the impact of xenon doping on contaminated liquid argon. The presence of xenon increased the amount of collected light, consistently across all photodetectors, indicating that argon-xenon interactions do supersede the impact of argon-nitrogen collisions that would otherwise quench the scintillation light production.

An increase in light production from the contaminated argon was observed at the lowest doping level tested (1.1~ppm) and continued to increase up to around 16 ppm, where the detected response appeared to flatten out at a level comparable to that previously obtained in pure argon. This behavior was observed in each of the four different types of photodetectors in the test. Another important result is that there was no change in the charge signal amplitude collected with the TPC throughout the doping operations.

We note that light output dependence on xenon concentration could be specific to our particular configuration and mixture of Ar-Xe-N$_2$. The ratio of the light collected by the two X-ARAPUCA sensors demonstrates that the excitation energy is indeed transferred from argon to xenon. Distinguishing between \textit{fast} and \textit{slow} component of the original argon light appears to confirm the hypothesis that the energy transfer happens on the meta-stable triplet state of argon excimer Ar$_2^*$; on the other hand, an unexpected drop in the argon \textit{fast} component is observed as soon as the first~ppm of xenon is introduced. The light output was observed to be stable during the measurement period of several weeks after the doping operations were concluded.

Studies of light attenuation along the TPC drift distance also confirm light recovery with respect to the period with nitrogen contamination, and they show a relative increase in the amount of light collected far away from the photosensors. This effect is attributed to the larger Rayleigh scattering length of the xenon-produced 178~nm photons in argon (with respect to 127~nm photons in pure argon): this should lead to a much more uniform light response in the DUNE far detectors, with respect to the undoped argon condition.

As mentioned in section~\ref{sec:analysis_xArapuca}, a similar campaign was performed in mid-2020 with the other DUNE prototype at CERN, ProtoDUNE Dual-Phase (DP). However, in addition to the substantially different photon detectors technology employed by the two DUNE prototypes, other aspects of the experimental set-ups were dissimilar. The DP TPC geometry allowed for drift lengths of up to 6~m, with the light detection system deployed in an array at the bottom of the cryostat beneath a partially transparent HV cathode. Furthermore, the datasets of muon tracks were selected with the distinct external and internal trigger systems used by the two detectors. 

Overall, the SP and DP detectors report similar qualitative behaviour of the scintillation light output, such as the depletion of the \textit{fast} component and the relative increase in collected light farther from the photon detectors. However, there are some quantitative differences in the results that will require more detailed analysis, in order to disentangle detector effects. A planned combined analysis of the ProtoDUNE SP and DP datasets will provide a more refined understanding the argon-xenon interaction mechanism.

In conclusion, the results obtained so far in ProtoDUNE-SP show that xenon doping is a valid technique to boost the photon detection performances of large drift TPCs, thus validating the plan to employ it for the second far detector module of DUNE~\cite{VDreview}.

\clearpage

\appendix
\section{Xenon injections in ProtoDUNE-SP and contamination}
\label{sec:appendix_inj}

A more in depth description of the actual xenon injection procedure in ProtoDUNE-SP is reported here. As mentioned in section~\ref{sec:xe_doping_schedule} with reference to the ProtoDUNE-SP gas re-circulation circuit, the xenon injection point is placed along the chimney boil-off re-circulation line, way before the argon condenser, in order to ensure full argon-xenon mixing within the gas flow.

\begin{figure}[ht]
    \centering 
     \includegraphics[width=.31\textwidth,]{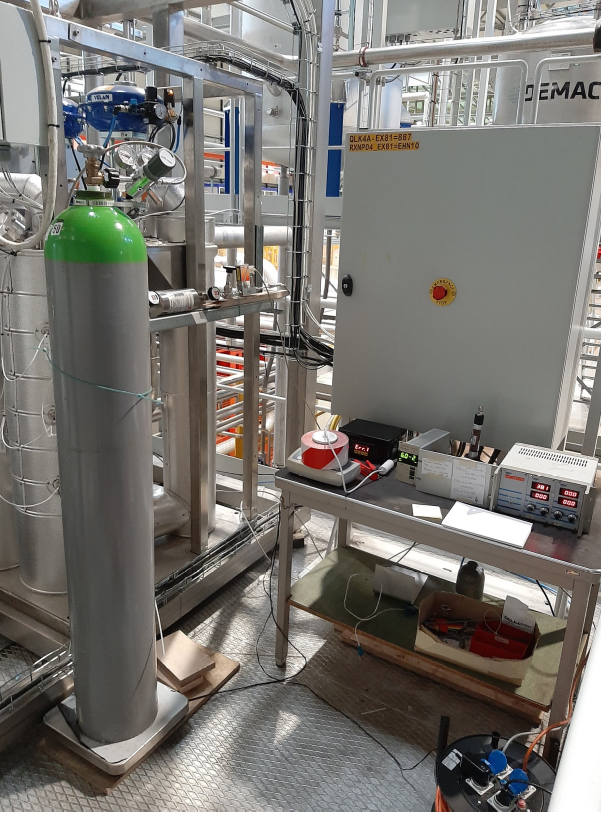}
     \includegraphics[width=.6\textwidth,]{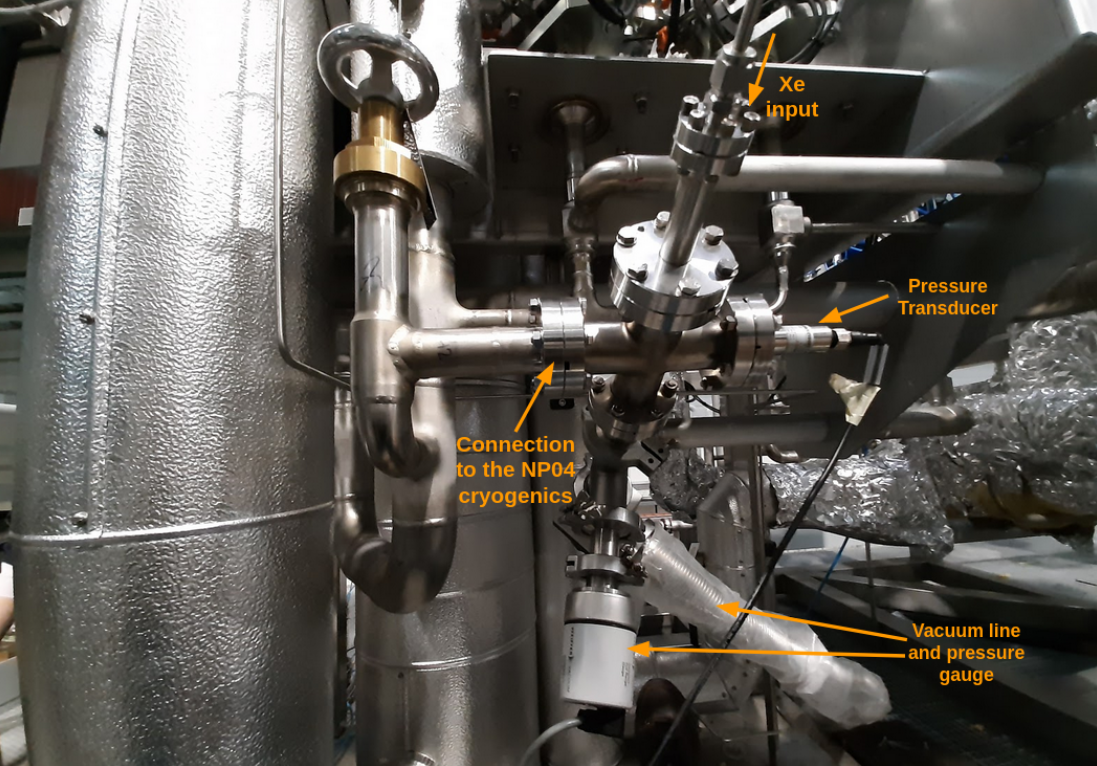}
    \caption{\label{fig:xe-injection-line}Left: the xenon bottle - on the scale - connected to the gas purifier, the mass flow-meter and the injection line. Right: detail of the UHV injection line equipped with vacuum/pressure monitoring devices and connected to the NP04 gas circulation system.}
\end{figure}

In order to precisely control the amount of gas introduced at any step of the doping, xenon bottles were placed on a scale connected to the detector slow control system.
A dedicated purification filter (SAES Micro-Torr\footnote{General product specification:\\ $http://www.saespuregas.com/Library/purifier\_specifications/902\_Media\_Specification.pdf$.}) was installed on the line, followed by a mass flow-meter, calibrated for xenon, and a pressure gauge. The entire line installed between the xenon bottle and the connection with the argon re-circulation system was kept under vacuum by a separate pumping system.
Xenon pressure, flow and bottle weight were continuously recorded by the slow control system. Figure~\ref{fig:xe-injection-line} illustrates the xenon injection set up.

\begin{figure}[ht]
    \centering 
    \includegraphics[width=.85\textwidth,]{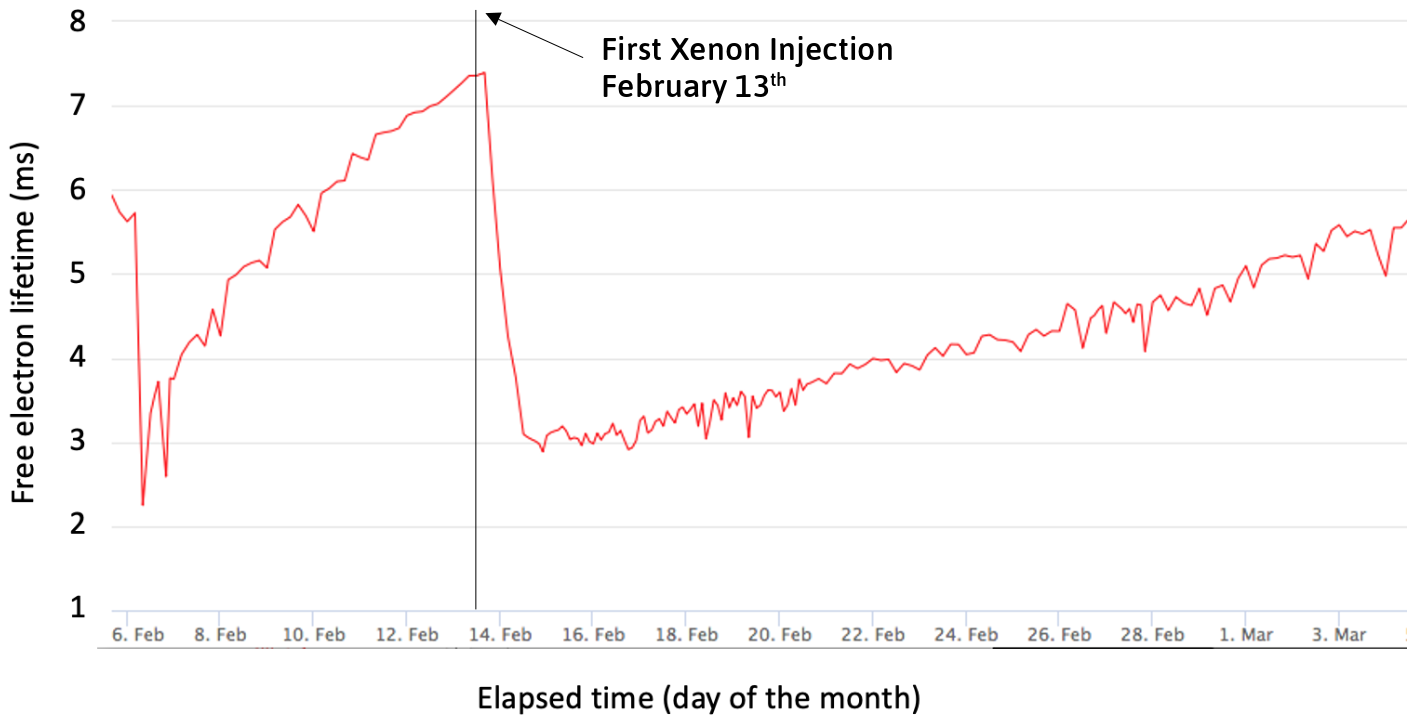}
    \caption{\label{fig:SF6-effect}Free electron lifetime measurement in ProtoDUNE-SP performed with dedicated purity monitors. The linear drop recorded around February $13^{\mathrm{th}}$ coincides with the first xenon injection and is attributed to the presence of fluorinated contaminants in the bottle. The subsequent recovery rate, due to LAr recirculation, is about a factor 8 to 10 slower than in previous recoveries (exemplified in the increase shown prior to the injection). This suggests that the ProtoDUNE purifiers can absorb fluorinated compounds, though with a factor~$\sim$10 lower efficiency, with respect to oxygen.}
\end{figure}

The doping was performed with three different bottles of xenon.
The first one (containing about 3~kg of gas) was rated with a purity grade 5.0\footnote{The purity grade refers to the fractional amount of gas in the bottle. 5.0 grade corresponds to 99.999\% of xenon in the bottle, or 10~ppm of contaminants overall.}, without any specifications on upper limits on fluorinated compounds. However, during the first injection, measurements with dedicated purity monitors highlighted a sizable degradation of the free electron lifetime within the LArTPC, as shown in figure~\ref{fig:SF6-effect}. The same effect was witnessed when turning on the TPC and recording a lower than usual amount of charge at the anode (as discussed for figure~\ref{fig:sigstrenxe}).

As a consequence, xenon injection was stopped and a set of spectrographic/chromatograpic analyses were performed at CERN~\cite{Corbetta_M}. Electro-negative impurities were identified as C$_2$F$_6$ ($\sim$10~ppm) plus traces of SF$_6$ and CO$_2$. These compounds, that can be present in xenon at the ppm level as residuals of the distillation process, are known to be highly electro-negative (several orders of magnitude higher than oxygen~\cite{Bakale_1976}), hence they can significantly degrade the free electron lifetime in LAr even at concentrations of few~ppt.
After this episode, free electron lifetime in ProtoDUNE-SP slowly recovered with a time constant of $\sim$30~days, indicating that the purifiers are able to absorb fluorinated compounds, albeit with an efficiency about 10~times lower than that for oxygen.

Two additional xenon bottles (containing about 17.5~kg each) were then acquired, rated with a purity grade of 5.5 and a specified SF$_6$ content certified by the producer to be lower than 20~ppb (following standard procedures set by CERN for the ATLAS and ALICE experiments). The higher-purity xenon produced no further sizable electron lifetime degradation in the TPC during the subsequent injections, and it allowed concluding the doping campaign successfully.

\clearpage

\centerline{\bf DUNE Acknowledgement}
\centerline{\bf November 2023}
\bigskip

The ProtoDUNE-SP detector was constructed and operated on the CERN Neutrino Platform. We gratefully acknowledge the support of the CERN management, and the CERN EP, BE, TE, EN and IT Departments for NP04/Proto\-DUNE-SP.

This document was prepared by the DUNE collaboration using the resources of the Fermi National Accelerator Laboratory (Fermilab), a U.S. Department of Energy, Office of Science, HEP User Facility. Fermilab is managed by Fermi Research Alliance, LLC (FRA), acting under Contract No. DE-AC02-07CH11359.

This work was supported by
CNPq,
FAPERJ,
FAPEG and 
FAPESP,                         Brazil;
CFI, 
IPP and 
NSERC,                          Canada;
CERN;
M\v{S}MT,                       Czech Republic;
ERDF, 
H2020-EU and 
MSCA,                           European Union;
CNRS/IN2P3 and
CEA,                            France;
INFN,                           Italy;
FCT,                            Portugal;
NRF,                            South Korea;
CAM, 
Fundaci\'{o}n ``La Caixa'',
Junta de Andaluc\'ia-FEDER,
MICINN, and
Xunta de Galicia,               Spain;
SERI and 
SNSF,                           Switzerland;
T\"UB\.ITAK,                    Turkey;
The Royal Society and 
UKRI/STFC,                      United Kingdom;
DOE and 
NSF,                            United States of America.

\bibliographystyle{JHEP}
\bibliography{bibliography} 

\end{document}